\documentclass[12pt]{article}
\usepackage{amsfonts}
\usepackage{amsmath}
\usepackage{latexsym}
\usepackage{amssymb}
\usepackage{epsfig}
\usepackage[latin1]{inputenc}
\usepackage{graphicx}
\usepackage{mathrsfs}

\newif{\ifcomentarios}
\comentariosfalse

\newtheorem{theorem}{Theorem}

\newtheorem{remark}[theorem]{Remark}
\newtheorem{lemma}[theorem]{Lemma}
\newtheorem{proposition}[theorem]{Proposition}

\newtheorem{conjecture}[theorem]{Conjecture}
\newtheorem{corollary}[theorem]{Corollary}

\newcommand{\dint}{\displaystyle\int}
\renewcommand{\mathcal}{\mathscr}
\renewcommand{\mathbf}{\boldsymbol}

\begin{document}

\title{On the Mayer series of two-dimensional Yukawa gas at inverse
temperature in the interval of collapse}
\author{{Wilhelm Kroschinsky \thanks{
Email: wilhelm.kroskinsque@usp.br} \quad and \quad Domingos H. U. Marchetti 
\thanks{
Email: marchett@if.usp.br}} \\
%EndAName
Instituto de Física\\
Universidade de São Paulo\\
05508-090 São Paulo, SP, Brazil}
\date{}
\maketitle

\begin{abstract}
A Theorem on the minimal specific energy for a system with $\pm 1$ charged
particles interacting through the Yukawa pair potential $v$ is proved which
may stated as follows. Let $v$ be represented by scale mixtures of
$d$--dimensional Euclid's hat (cutoff at short scale distances) with
$d\geq 2$. 
For any even number of particles $n$, the interacting energy $U_{n}$ divided
by $n$, attains an $n$--independent minimum at a configuration with zero net
charge and particle positions collapsed altogether to a point. For any odd
number of particles $n$, the ratio $U_{n}/(n-1)$ attains its minimum value,
the same of the previous cases, at the configuration with $\pm 1$ net charge
and particle positions collapsed to a point. This Theorem is then used to
resolve an obstructive remark of an umpublished paper (Remark 7.5 of
\cite{Guidi-Marchetti}) which, whether the standard decomposition of
the Yukawa 
potential into scales were adopted, would impede a direct proof of the
convergence of the Mayer series of the two-dimensional Yukawa gas for the
inverse temperature in the whole interval $[4\pi ,8\pi )$ of collapse. In
the present paper, it is proven convergence up to the second threshold $6\pi 
$ and its given explanations on the mechanism that allow it to be extend up
to $8\pi $. The paper distinguishes the matters concerning stability from
those related to convergence of the Mayer series. In respect to the latter
the paper dedicates to the Cauchy majorante method applied to the density
function of Yukawa gas in the interval of collapses. It also dedicates to
the proof of the main Theorem and estimates of the modified Bessel functions
of second kind involved in both representations of two-dimensional Yukawa
potential: standard and scale mixture of the Euclid's hat function.

\noindent \textbf{MSC:} 60J45, 42A82, 97K10, 82B21, 82B28

\noindent \textbf{Keywords:} two-dimensional Yukawa potential, minimal specific
energy, scale mixtures of Euclid's hat, stability, Mayer series,
collapse interval
\end{abstract}

\section{Introduction and background of tools and methods}

\setcounter{equation}{0} \setcounter{theorem}{0}

The present paper investigates a system of particles with $\pm 1$ charges
living in a two dimensional Euclidian space and interacting through the
Yukawa pair potential $v(x)=\left( -\Delta +1\right) ^{-1}(0,x)$. Because of
Yukawa and Coulomb potentials look like the same at short distances, the
two-dimensional Yukawa gas inherits the same instabilities of the
corresponding Coulomb system when the inverse temperatures $\beta $ belongs
to the interval $[4\pi ,8\pi )$, in which a sequence of collapses of neutral
cluster of size $2n$ occurs at the thresholds $\beta _{2n}=8\pi \left(
1-1/2n\right) $, $n\in \mathbb{N}$. It remains an open problem for this
system to establish convergence of the Mayer series in powers of activity $z$,
with the first even terms removed from the series, how many depending on $
\beta \in \lbrack 4\pi ,8\pi )$. It is our purpose to revisit this long
standing problem.

Benfatto\cite{Benfatto} and collaborators from the Italian school (see
references therein) initiate a program using iterated Mayer series for
pressure (and correlation functions) together with ideas from the work of
Gopfert-Mack \cite{Gopfert-Mack} and Imbrie \cite{Imbrie}. Brydges and
Kennedy \cite{Brydges-Kennedy} have also considered the Mayer expansion of
the two-dimensional Yukawa gas in the context of the Hamilton-Jacobi
equation. We adopt in present investigation their continuum scaling
renormalization method, adding to that approach a new ingredient. The
novelty is related with the (short--range) decomposition of the Yukawa
potential into scales. Instead of the standard decomposition $v(x)=%
\displaystyle\int_{-\infty }^{0}\left( d\left( -\Delta +e^{-s}\right)
^{-1}(0,x)/ds\right) ds$ (or the discrete version of it) adopted in the
previous work, we use the scale mixture $v(x)=\displaystyle%
\int_{0}^{1}g(s)h(\left\vert x\right\vert /s)ds$ of Euclid's hat $h(r)$.
Using a concept introduced by Basuev \cite{Basuev1}, we first prove a
theorem that the minimal specific energy $e(v)$ and the constrained (to non
zero net charge) modified minimal specific energy $\bar{e}(v)$ are equal.

Our main theorem on specific energy when applied in the investigation of the
two-dimensional Yukawa gas resolve an obstructive limitation that has been
posed in an umpublished paper by Guidi and one of the authors (see
Conjecture 2.3 and Remark 7.5 of \cite{Guidi-Marchetti}) towards a direct
proof of the convergence of the Mayer series on the entire interval $[4\pi
,8\pi )$ whether the standard decomposition of the Yukawa potential were
adopted. The limitation value for the $3$--particles interacting energy
given by a numerical evaluation in \cite{Guidi-Marchetti} is proven in
Proposition \ref{bare3} and Remark \ref{e3estimate}.

In the present paper, the convergence of the Mayer series is proven up to
the second threshold $\beta \in \lbrack 4\pi ,6\pi )$ and it is provided a
full explanation of the mechanisms that allow it to be extend up to $8\pi $.
We distinguished the issues concerning stability from those related to
convergence of the Mayer series. In respect to the latter, the paper
dedicates in Section \ref{MDF} to the Cauchy majorante method applied to the
density function of Yukawa gas in the region of collapse. Regarding the
former, we considered only the simplest case of collapse prevention of
neutral pair of charges due the presence of other charges in the
configuration.

We shall now review the tools and methods employed in present investigation.
We refer to the references for detail.

\paragraph{Decomposition of radial positive functions of positive type.}

Positive definite functions have arisen in many areas of (pure and applied)
mathematics and physics (see \cite{Stewart} for an historical survey). A
continuous function $f$ defined in $\mathbb{R}^{d}$ is called positive
definite (abbreviated as p. d.) if the $n\times n$ real matrix $\left[
f\left( x_{i}-x_{j}\right) \right] _{1\leq i,j\leq n}$ is positive definite
for $n\in \mathbb{N}$ arbitrary elements $x_{1}$, \ldots , $x_{n}$ of $%
\mathbb{R}^{d}$:%
\begin{equation}
\sum_{1\leq i,j\leq n}\bar{z}_{i}z_{j}f\left( x_{i}-x_{j}\right) \geq
0~,\qquad \forall z_{1},\ldots ,z_{n}\in \mathbb{C}~.  \label{pd}
\end{equation}%
The celebrate work of Bochner (see e.g \cite{Bochner}) characterizes these
functions as follows: $f$ (with $f(0)=1$) is positive definite if, and only
if, is a Fourier-Stieltjes transform $\check{\mu}(x)=\displaystyle\int_{ 
\mathbb{R}^{d}}e^{ix\cdot \xi }d\mu (\xi )$ of a probability Borel measure $
\mu $ on $\mathbb{R}^{d}$. Although powerful, Bochner's theorem may be
difficult to use in practice: how do we know that a given $f$
satisfies (\ref{pd})? Even when explicit computation of Fourier
transform is available, how 
do we represent $f$ into suitable scale mixture of elementary functions?

Recently (see \cite{Gneiting, Hainzl-Seiringer, Jaming-Matolcsi-Revesz} and
references therein), investigations towards extending Bochner's theorem seek
for concrete examples and easy checkable criteria of p. d. function. A
particularly interesting subclass of p. d. functions, denoted in
\cite{Jaming-Matolcsi-Revesz} by $\Omega _{d}^{+}$, is provided by
radial continuous functions: $f(x)=\varphi (\left\vert x\right\vert )$
for some positive continuous 
function $\varphi $ of $\mathbb{R}_{+}$. A simple example of these functions
that vanishes out of a ball $B_{s}$ of radius $s>0$ centered at origin
is given by the Euclid's hat ($d=2$) 
\begin{equation*}
\frac{4}{\pi s^{2}}\chi _{s/2}\ast \chi _{s/2}(x)\equiv h(\left\vert
x\right\vert /s)
\end{equation*}%
where $\chi _{r}(x)=\chi _{B_{r}}(x)$ is the characteristic function of $
B_{r}$. In \cite{Jaming-Matolcsi-Revesz} Jaming, Matolcsi and Révész have
identified certain compactly supported functions, alike this one, as extrema
rays of the cone $\Omega _{d}^{+}$, playing the same role as the family $
\left\{ e^{i\xi \cdot x}\right\} $ for the Bochner's theorem. So, if $
\varphi $ is an extremum ray of $\Omega _{d}^{+}$ then, by Choquet
representation, 
\begin{equation}
\int_{0}^{\infty }\varphi \left( \left\vert x\right\vert /s\right) d\nu (s)
\label{choquet}
\end{equation}
is an element of $\Omega _{d}^{+}$ for a suitable positive measure $\nu $
supported on the family of scales $\left\{ \varphi \left( \left\vert
x\right\vert /s\right) \right\} $ of $\varphi $. An open problem is to find
all extrema of $\Omega _{d}^{+}$ (see \cite{Jaming-Matolcsi-Revesz}).

Geiting \cite{Gneiting} and Hainzl-Seiringer \cite{Hainzl-Seiringer} give,
on the other hand, complete characterizations of the subclass $H_{d}\subset
\Omega _{d}^{+}$ that are formed by scaling mixtures of $d$--dimensional
Euclid's hat, extending Polya's criterion on $\mathbb{R}^{d}$, for $d\geq 2$.
Hainzl-Seiringer's representation however suffices to make our point in
the present work. Let us start with the two-dimensional Yukawa potential,
which is an element of $\Omega _{2}^{+}$ given by the Green's function $v(1/
\sqrt{\kappa },x)=\left( -\Delta +\kappa \right) ^{-1}(0,x)$ (the resolvent
kernel of the Laplacian operator $\Delta =\partial ^{2}/\partial
x_{1}^{2}+\partial ^{2}/\partial x_{2}^{2}$). Applying Fourier transform
yields (with $v(x)\equiv v(1,x)$ and $v(1/\sqrt{\kappa },x)=v(1,\sqrt{\kappa 
}x)$. See e.g. Sec. 7.2 of \cite{Glimm-Jaffe}) 
\begin{equation}
v(x)=\frac{1}{2\pi }\int_{\mathbb{R}^{2}}e^{i\xi \cdot x}\frac{1}{\xi ^{2}+1}
d\xi =\frac{1}{2\pi }K_{0}(\left\vert x\right\vert )  \label{v}
\end{equation}
where $K_{0}$ is the modified Bessel function of second kind of order $0$.
Hainzl-Seiringer's formula for this function reads 
\begin{equation}
v(x)=\int_{0}^{\infty }h(\left\vert x\right\vert /s)g(s)ds  \label{vgh}
\end{equation}
where ($h(0)=1$) 
\begin{equation}
h(w)=\frac{2}{\pi }\left( \arccos w-w\sqrt{1-w^{2}}\right) ~,\qquad \text{if}
\ \quad 0<w\leq 1  \label{h}
\end{equation}
$h(w)=0$ if $w>1$ and 
\begin{equation}
g(s)=\frac{-s}{4\pi }\int_{s}^{\infty }K_{0}^{\prime \prime \prime }(r)\frac{
r}{\sqrt{r^{2}-s^{2}}}dr~.  \label{g}
\end{equation}
The mixture density function $g(s)$ for the Yukawa potential in $d=1$ and $3$
dimensions and the Coulomb potential in $d$--dimensions have closed forms
(see Examples 1 and 2 of \cite{Hainzl-Seiringer}). For the Yukawa function
in $2$--dimensions, however, $g(s)$ can only be written in term of Meijer $G$
--functions (see \cite{Beals-Szmigielki} for an introduction): $g(s)=\sqrt{%
\pi }G_{13}^{30}\left( s^{2}/4\left\vert
  \genfrac{}{}{0pt}{1}{1/2}{0,1,3}
\right. \right) /(2\pi s)$.

We observe that $h(w)$ is a convex function of $w\in \mathbb{R}_{+}$ and a
mixture of the Euclid's hat (\ref{vgh}) preserves convexity. This useful
property, as we shall see, distinguishes (\ref{vgh}) from another common
decomposition of (\ref{v}) into scales (see Fig. \ref{fig4}): with
$v(s,x)=\left( -\Delta
+1/s^{2}\right) ^{-1}(0,x)=\left( -\Delta +1\right)
^{-1}(0,x/s)=K_{0}(\left\vert x\right\vert /s)$, we write 
\begin{equation*}
v(x)=\int_{0}^{1}\dot{v}(s;x)ds~
\end{equation*}
by the fundamental theorem of calculus. Substituting the derivative $\dot{v}%
(s;x)$ with respect to $s$, yields 
\begin{equation}
v(x)=\frac{1}{2\pi }\int_{0}^{1}\tilde{h}(\left\vert x\right\vert /s)\frac{ds%
}{s}  \label{v1}
\end{equation}%
where $\tilde{h}(w)=-wK_{0}^{\prime }(w)=wK_{1}(w)$, with $K_{1}$ the
modified Bessel function of second kind of order $1$, like $h$ given
by (\ref{h}), decreases monotonously to $0$ and satisfies $h_{1}(0)=1$
but changes 
from concave to convex as $w\in \mathbb{R}_{+}$ varies. The mixture density $
g(s)$ for both decompositions of $v$, (\ref{vgh}) and (\ref{v1}), behaves in
the neighborhood of $s=0$ as $\left( 2\pi s\right) ^{-1}$ implying that $%
v(x) $ behaves as the Coulomb potential $(-1/2\pi )\log \left\vert
x\right\vert $ at short distances.

\paragraph{Gaussian Processes and renormalization group.}

Positive definite functions plays an important role on renormalization group
(RG) methods in statistical physics. Brydges and collaborators
\cite{Brydges-Guadagni-Mitter} (see also \cite{Brydges-Talarczyk})
coined a term ``finite range decomposition'' to the mixture of different
scales (\ref{choquet}) for some compactly supported radial extremal
functions $\varphi $. They used a probabilistic argument as follows:
breaking up the range of 
integration into disjoint union of intervals $I_{j}=[L^{-j},L^{-j+1})$, $
j\geq 1$ for $L>1$ and $I_{0}=[1,\infty )$, (\ref{vgh}) may be seen as the
\textquotedblleft finite range\textquotedblright\ decomposition 
\begin{equation}
\phi =\sum_{j\geq 0}\zeta _{j}  \label{phi}
\end{equation}%
of a Gaussian process $\phi $ of mean $\mathbb{E}\phi (x)=0$ and covariance $
\mathbb{E}\phi (x)\phi (y)=v\left( x-y\right) $ into a family of independent
Gaussian processes $\left\{ \zeta _{j}\right\} $ of mean $\mathbb{E}\zeta
_{j}=0$ and covariance 
\begin{equation*}
\mathbb{E}\zeta _{j}(x)\zeta _{j}(y)=\int_{I_{j}}g(s)h\left( \left\vert
x-y\right\vert /s\right) ds\equiv v_{I_{j}}(x-y).
\end{equation*}%
Since the covariance of a sum of independent Gaussian random variable is the
sum of their covariances, $v=\displaystyle\sum_{j\geq 0}v_{I_{j}}$
equals (\ref{vgh}). The authors of \cite{Brydges-Guadagni-Mitter,
  Brydges-Talarczyk} 
were also capable of applying suitable finite range decomposition to a large
class of positive definite functions on $\mathbb{R}^{d}$ and $\mathbb{Z}^{d}$
that comprises integral kernels (Green's functions) of certain elliptic
operators, their corresponding finite differences and fractional powers.

When a statistical system is represented by the expectation $\mathbb{E}
\mathcal{Z}$ of a functional $\mathcal{Z}\left( \phi \right) $, the
decomposition (\ref{phi}) of the Gaussian field $\phi $ can be used to
integrate out each $\zeta _{j}$ at a time. Let $\mathbb{E}^{(j)}$ denote the
expectation with respect the Gaussian field $\zeta _{j}$. The
renormalization group is a method of calculating the expectation $\mathbb{E}
\mathcal{Z}$ through the sequence of maps $\mathcal{Z}_{j}\longmapsto 
\mathcal{Z}_{j+1}=\mathbb{E}^{(j+1)}\mathcal{Z}_{j}$ starting from $\mathcal{
Z}_{0}=\mathcal{Z}$. The limit $\lim_{j\rightarrow \infty }\mathcal{Z}_{j}=
\mathbb{E}\mathcal{Z}$, supposing it exists, is obtained provided $\mathcal{Z
}_{j}\longmapsto \mathcal{Z}_{j+1}$ is amenable to be analyzed as a
dynamical system depending on parameters in the initial condition. For
instance, in the decomposition (\ref{phi}) of $\phi $ into finite range
fields $\zeta _{j}$ corresponding to the Yukawa potential (\ref{vgh}), the
limit $j\rightarrow \infty $ drives the statistical system into the short
scaling limit $s\rightarrow 0$ for which the potential diverges
logarithmically. For an infinitely many-particle system with $\pm 1$
charges, the existence of $\lim_{j\rightarrow \infty }\mathcal{Z}_{j}$
expresses the thermodynamical stability of the system. We shall come back to
this issue below.

\paragraph{Hamilton--Jacobi equation and majorant method.}

Under the Kac--Siegert transformation, \linebreak \cite{Frohlich-Spencer,
Brydges-Martin} the grand partition function for the two--dimensional Yukawa
gas of particles with $\pm 1$ charges can be written as the expectation $
\mathbb{E}\mathcal{Z}_{0}$ (with respect to the Gaussian field $\phi $) of 
\begin{eqnarray}
\mathcal{Z}_{0}\left( \phi \right)  &=&\exp \left( \mathcal{V}_{0}(\phi
)\right) \,,  \notag \\
\mathcal{V}_{0}(\phi ) &=&z\int_{\mathbb{R}^{2}}:\cos \sqrt{\beta }\phi
(x):_{v}dx  \notag \\
&=&\sum_{\sigma \in \left\{ -1,1\right\} }\int_{\mathbb{R}^{2}}dx~z:e^{i%
\sqrt{\beta }\sigma \phi (x)}:_{v}  \label{kac}
\end{eqnarray}%
where the parameters $\beta $ and $z$ are, respectively, the inverse
temperature and activity and $:\cdot :_{v}$ indicates Wick ordering with
respect to the potential $v$. In the present work, we shall adopt the
continuum scale decomposition (\ref{v}) instead of (\ref{phi}). The induced
RG dynamics is thus generated by a Hamilton-Jacobi equation as proposed in 
\cite{Brydges-Kennedy} by Brydges and Kennedy. Let us expand these ideas in
some detail. A scale--dependent--interaction $v:\mathbb{R}_{+}\times \mathbb{
R}^{2}\longrightarrow \mathbb{R}$ is introduced replacing (\ref{vgh}) by a
mixture supported in a finite interval $\left[ t_{0},t\right] $ of scales 
\begin{equation}
v(t,x)=\int_{t_{0}}^{t}h(\left\vert x\right\vert /s)g(s)ds  \label{vtx}
\end{equation}%
where $t_{0}>0$ is a cutoff of the short scale distances. Since $g$ and $h$
are continuous, we have $\lim_{t\searrow t_{0}}v(t,x)\equiv 0$. The
renormalization group is now given by a convolution mapping $\left( t,\phi
\right) \longmapsto \mathcal{Z}(t,\phi )=\mathbb{E}^{(t)}\mathcal{Z}_{0}
\mathcal{(\phi +\cdot )}$ with initial data $\mathcal{Z}(t_{0},\phi )=%
\mathcal{Z}_{0}\mathcal{(\phi )}$, where $\mathbb{E}^{(t)}$ denotes the
expectation with respect the Gaussian field $\zeta $ with covariance $
v\left( t,x-y\right) $. Formally, $\mathcal{Z}(t,\phi )$ satisfies the
initial value problem of a \textquotedblleft heat equation\textquotedblright 
\begin{equation*}
\frac{\partial \mathcal{Z}}{\partial t}=\frac{1}{2}\Delta _{\dot{v}}\mathcal{
\ Z}~~,\qquad \lim_{t\searrow t_{0}}\mathcal{Z}(t,\phi )=\mathcal{Z}_{0}
\mathcal{(\phi )}
\end{equation*}
where $\dot{v}(t,x):=\partial v/\partial t(t,x)=g(t)h(\left\vert
x\right\vert /t)$, by the fundamental theorem of calculus, is the weighted
Euclid's hat scaled by $t$ and $\Delta _{\dot{v}}$ is the \textquotedblleft
Laplacian\textquotedblright\ operator 
\begin{equation}
\Delta _{\dot{v}}\mathcal{Z=}\int_{\mathbb{R}^{2}\times \mathbb{R}^{2}}dxdy
\dot{v}(t,x-y)\frac{\delta ^{2}\mathcal{Z}}{\delta \phi (x)\delta \phi (y)}
~~.  \label{Delta}
\end{equation}
Writing $\mathcal{Z}(t,\phi )=\exp \left( \mathcal{V}(t,\phi )\right) $, the
heat equation turns into a nonlinear equation for $\mathcal{V}$: 
\begin{equation}
\frac{\partial \mathcal{V}}{\partial t}=\frac{1}{2}\Delta _{\dot{v}}\mathcal{
\ V+}\frac{1}{2}\left( \nabla \mathcal{V},\nabla \mathcal{V}\right) _{\dot{v}
}~~,\qquad \lim_{t\searrow t_{0}}\mathcal{V}(t,\phi )=\mathcal{V}_{0}
\mathcal{(\phi )}  \label{Veq}
\end{equation}
where $\Delta _{\dot{v}}$ acts as in (\ref{Delta}) and 
\begin{equation}
\left( \nabla \mathcal{V},\nabla \mathcal{V}\right) _{v}=\int_{\mathbb{R}
^{2}\times \mathbb{R}^{2}}dxdy\dot{v}(t,x-y)\frac{\delta \mathcal{V}}{\delta
\phi (x)}\frac{\delta \mathcal{V}}{\delta \phi (y)}~.  \label{grad2}
\end{equation}

In \cite{Brydges-Kennedy}, the authors considered the random field $\phi $
on $\mathbb{Z}^{d}$ instead, for which the functional derivative $
\displaystyle\int dx~v(x)~\delta /\delta \phi (x)~\mathcal{V}(\phi
)=\lim_{\varepsilon \rightarrow 0}\left( \mathcal{V}(\phi +\varepsilon v)-
\mathcal{V}(\phi )\right) /\varepsilon $ becomes partial derivative $%
\partial /\partial \phi _{x}$ with respect to the variable $\phi _{x}\in 
\mathbb{R}$ at site $x$. Because of translation invariance, (\ref{Delta})
and (\ref{grad2}) diverges even if $\mathbb{R}^{2}$ is replaced by $\mathbb{Z
}^{2}$ but this can be solved by fixing one point $x$ of $\mathbb{Z}^{2}$.
Inserting the Taylor expansion (multi-index formula): 
\begin{equation*}
\mathcal{V}\left( t,\phi \right) =\sum_{n\geq 1}\sum_{\alpha :\left\vert
\alpha \right\vert =n}\frac{1}{\alpha !}\frac{\partial ^{\alpha }\mathcal{V}
}{\partial \phi ^{\alpha }}(t,0)~\phi ^{\alpha }
\end{equation*}
into an integral equation equivalent to (\ref{Veq}), a system of equations
for derivatives of $\mathcal{V}$ (by collecting order by order terms),
together with an appropriate norm, is used to majorize $\mathcal{V}\left(
t,\phi \right) $ by the solution $\nu (t,\varphi )$ of a first order PDE
equation in two independent real variables $\left( t,\varphi \right) $, $
\varphi $ playing the role of chemical potential. The local existence and
uniqueness of the initial value problem (\ref{Veq}) are then proved in ref. 
\cite{Brydges-Kennedy} (see Theorem 2.2 and Proposition 2.6 therein) for a
domain in plane $\left( t,z\right) $ with $z=e^{\varphi }$ ($\beta $ may be
included as well). Quoting the authors, these results are \textquotedblleft
the precise version of the Mayer expansion\textquotedblright\ for the
pressure or correlations functions of statistical systems.

Brydges and Kennedy have also provided an equivalent system of ordinary
differential equations for the Ursell functions (Lemma 3.3 of
\cite{Brydges-Kennedy}) which replaces (\ref{Veq}) defined on
$\mathbb{Z}^{d}$ 
and can be used for systems of point particles in $\mathbb{R}^{d}$. If $
\left( \Omega ,\mathcal{B},d\varrho (\zeta )\right) $ denotes the finite
measure space on $\left\{ -1,1\right\} \times \mathbb{R}^{2}$ corresponding
to the possible states of a single particle (we united $\sigma $ and $x$
into $\zeta =\left( \sigma ,x\right) $), the solution of (\ref{Veq}) may be
represented formally as 
\begin{equation}
\mathcal{V}\left( t,\phi \right) =\sum_{n\geq 1}\frac{1}{n!}\int
d^{n}\varrho \psi _{n}^{c}(t,\zeta _{1},\ldots ,\zeta _{n}):\!\exp \left( i
\sqrt{\beta }\displaystyle\sum\nolimits_{j=1}^{n}\sigma _{j}\phi
(x_{j})\right) \!:  \label{V}
\end{equation}%
where the Ursell functions $\psi _{n}^{c}(t,\zeta _{1},\ldots ,\zeta _{n})$
are translational invariant and invariant under the action of the symmetric
group $\mathbb{S}_{n}$ of permutations of the index set $\left\{ 1,\ldots
,n\right\} $.\footnote{
Applying the functional calculus on (\ref{V}) we obtain formally from
(\ref{Veq}) the system of equations (see eq. (\ref{systeqs}))
satisfied by the $ 
\psi _{n}^{c}$'s. For instance, the Laplacian of $\mathcal{V}$ gives 
\begin{eqnarray*}
\Delta _{\dot{v}}\mathcal{V}\left( t,\phi \right)  &=&\lim_{\varepsilon
,\eta \rightarrow 0}\frac{1}{\varepsilon \eta }\left( \mathcal{V}\left(
t,\phi +\varepsilon \dot{v}+\eta \dot{v}\right) -\mathcal{V}\left( t,\phi
+\varepsilon \dot{v}\right) -\mathcal{V}\left( t,\phi +\eta \dot{v}\right) +
\mathcal{V}\left( t,\phi \right) \right)  \\
&=&\sum_{n\geq 1}\frac{1}{n!}\int d^{n}\varrho \frac{-\beta }{2}\sum_{i\neq
j}\sigma _{i}\sigma _{j}\dot{v}(t,x_{i}-x_{j})\psi _{n}^{c}(t,\zeta
_{1},\ldots ,\zeta _{n}):\!\exp \left( i\sqrt{\beta }\sum\nolimits_{j=1}^{n}
\sigma _{j}\phi (x_{j})\right) \!:\ .
\end{eqnarray*}%
.} To make mathematical sense of the above equations (\ref{Veq}) and
(\ref{V}) one can check, at the very end, whether the solution of the
system of 
ODE's for $\psi _{n}^{c}$'s agrees with the statements of \cite{Lenard1,
Lenard2} on point processes of infinitely many particles (consult
\cite{Ruelle} for the definition of $n$--point correlation and cluster
functions 
and Theorem 5.4 of \cite{Gielerak} for a hybrid approach combining methods
employed for Poisson point process with correlation functions satisfied by
the (sine-Gordon) representation (\ref{kac}) of the Yukawa gas). The present
work will take the system of equations satisfied by the $\psi
_{n}^{c}(t,\zeta _{1},\ldots ,\zeta _{n})$ (see (\ref{systeqs}) and
(\ref{integralf}) below), together with the scale decomposition
(\ref{vtx}) for the Yukawa potential, as the starting point for our analysis.

\paragraph{Stability condition and minimal specific energy.}

Stability of the interaction $v$ is a condition under which there exist
the thermodynamic functions describing an infinitely large statistical
system. Let $U_{n}$ be the total energy potential of the classical charged
system of $n$ point particles at positions $x_{1}$, \ldots , $x_{n}$ of $
\mathbb{R}^{2}$, with respective charges $\sigma _{1}$, \ldots , $\sigma
_{n}\in \left\{ -1,1\right\} $, interacting through a pair Yukawa potential: 
\begin{equation}
U_{n}(\zeta _{1},\ldots ,\zeta _{n};v)=\sum_{1\leq i<j\leq n}\sigma
_{i}v\left( x_{i}-x_{j}\right) \sigma _{j}\ .~  \label{Unzetas}
\end{equation}
An interacting potential $v$ satisfies the \textit{stability condition} if
there exists $B>0$ such that 
\begin{equation}
U_{n}(\zeta _{1},\ldots ,\zeta _{n};v)\geq -nB  \label{stability}
\end{equation}
holds for all $(\zeta _{1},\ldots ,\zeta _{n})$ on the configurations space $
\displaystyle\bigcup_{n}$ $\left( \left\{ -1,1\right\} \times \mathbb{R}
^{2}\right) ^{n}$ (otherwise the specific energy $U_{n}/n$ would not be
bounded from below).

The standard stability theorem for charged system due to Fisher and Ruelle 
\cite{Fisher-Ruelle} (see Theorem I and eq. (III.7) therein) assures that:
if $\hat{v}(\xi )=\left( 1/2\pi \right) \int_{\mathbb{R}^{2}}v(x)e^{-i\xi
\cdot x}dx\geq 0$ and $v(0)=\left( 1/2\pi \right) \int_{\mathbb{R}^{2}}\hat{v
}(\xi )d\xi <\infty $, then
\begin{equation}
U_{n}(\zeta _{1},\ldots ,\zeta _{n};v)\geq -\frac{1}{2}v(0)\sum_{j=1}^{n}
\sigma _{j}^{2}  \label{Un}
\end{equation}
and, since $\sigma _{j}^{2}=1$, (\ref{stability}) is satisfied with $
B=v(0)/2 $. The proof of (\ref{Un}) follows from the \textquotedblleft
if\textquotedblright\ direction of Bochner's theorem. For this, note that
adding $1/2$ of each $i=j$ diagonal terms to (\ref{Unzetas}) (i.e.,
(\ref{Un}) with the right hand side passed to the left), the quadratic
form has to be 
positive as $v$ is positive definite. It follows from (\ref{v}) that $\hat{v}
(\xi )=\left( 2\pi \left( \xi ^{2}+1\right) \right) ^{-1}\geq 0$ is a
positive density but $v(x)$, which yields the stability constant $B$, grows
unboundedly at $x=0$. As the self--energy $v(0)$ diverges logarithmically,
the decomposition (\ref{v1}) or (\ref{vgh}) has to be used instead. The
thermodynamic functions are defined when the scales smaller than an $s_{0}>0$
are removed from the decomposition of $v$, but one has to prove that they
remain well defined after the cutoff is removed.

Let $v$ be the scale mixtures of Euclid's hat (\ref{vtx}), cutoff on the
short scales. We introduce the minimal specific energy $e=e(h)$ of $h$ at
the scale $s=1$ 
\begin{eqnarray}
e &=&\inf_{n\geq 2}e_{n}  \notag \\
e_{n} &=&\inf_{\left( \zeta _{1},\ldots ,\zeta _{n}\right) }\frac{1}{n}%
U_{n}(\zeta _{1},\ldots ,\zeta _{n};h)  \label{e}
\end{eqnarray}%
and the modified minimal specific energy $\bar{e}=\bar{e}(h)$, 
\begin{equation}
\bar{e}=\inf_{n\geq 2}\inf_{\substack{ \left( \zeta _{1},\ldots ,\zeta
_{n}\right)  \\ \mathrm{non-neutral}}}\frac{1}{n-1}U_{n}(\zeta _{1},\ldots
,\zeta _{n};h)  \label{ebar}
\end{equation}%
where the infimum is now taken over all non--neutral configurations $\left(
\zeta _{1},\ldots ,\zeta _{n}\right) $: $\left( x_{1},\ldots ,x_{n}\right)
\in \mathbb{R}^{2n}$ and $\left( \sigma _{1},\ldots ,\sigma _{n}\right) \in
\left\{ -1,1\right\} ^{n}$ such that $\sum_{j=1}^{n}\sigma _{j}\neq 0$. It
is clear that minimal specific energy (modified or not) of the scaled
Euclid's hat $h(\cdot /s)$ with $s\neq 1$ have all the same value by
homogeneity of the infimum and we have 
\begin{equation}
e\left( v\right) =\int_{t_{0}}^{t}e\left( h(\cdot /s)\right) g(s)ds=e\left(
h\right) \cdot \int_{t_{0}}^{t}g(s)ds\ .  \label{ev}
\end{equation}%
In the present paper we determine both specific energies $e$ and $\bar{e}$
and characterize the configuration that they are attained for $h$ and,
consequently, for (\ref{vtx}) by (\ref{ev}). From definitions
(\ref{stability}) and (\ref{e}), we have $-e(h)\leq h(0)/2=1/2$. We
show that 
this is in fact an equality and, moreover, $e=\bar{e}=-1/2$. More precisely,
we have proven in Sec. \ref{MSE} an improvement of (\ref{Un}) 
\begin{equation}
U_{n}(\zeta _{1},\ldots ,\zeta _{n};h)\geq \frac{1}{2}\left( \left\vert
\sum\nolimits_{j=1}^{n}\sigma _{j}\right\vert -\sum\nolimits_{j=1}^{n}\sigma
_{j}^{2}\right)  \label{UnSS}
\end{equation}%
from which the equality of specific energies follows at once. Observe that
the inequality (\ref{UnSS}) turns out to be an equality for certain
configurations.

We should mention a short note written by Basuev \cite{Basuev1} on the
minimal specific energy for classical one--specie system of particles in $
\mathbb{R}^{3}$, interacting through a radial two-body potential $\phi
(\left\vert x-y\right\vert )$. The conclusions of this investigation may be
stated as follows. Suppose that $\phi $ satisfies the two conditions that
defines what now-a-days is called Basuev
potentials:\cite{Lima-Procacci-Yuhjtman} there is $a>0$ such that:
\textbf{i. }$\phi \left( 
\left\vert x\right\vert \right) \geq \phi \left( a\right) >0$, for all $
\left\vert x\right\vert \leq a$; and \textbf{ii.} $\phi \left( a\right)
>2\mu (a)$ where
\begin{equation*}
\mu (a)=\sup_{n\geq 2}\sup_{\substack{ \left( x_{1},\ldots ,x_{n}\right) \in 
\mathbb{R}^{n}:  \\ \left\vert x_{i}-x_{j}\right\vert >a}}\sum_{i=1}^{n}\max
\left( -\phi \left( \left\vert x_{i}\right\vert \right) ,0\right)
\end{equation*}%
is finite (the supremum is taken over all configurations whose distance
between any pair exceeds $a$). Then $\phi $ and the potential $\phi ^{a}$,
given by $\phi ^{a}\left( \left\vert x\right\vert \right) =\phi \left(
\left\vert x\right\vert \right) $ if $\left\vert x\right\vert >a$ and $\phi
(\left\vert x\right\vert )=\phi (a)$ if $\left\vert x\right\vert \leq a$,
are stable with stability constant $B=\mu (a)/2$, and their minimal specific
energy are equal: $e\left( \phi \right) =e\left( \phi ^{a}\right) $ and $%
\bar{e}\left( \phi \right) =\bar{e}\left( \phi ^{a}\right) $. Colloquially,
it says that an increase of the positive part of the potential does not
reduce the binding energy of the system. Basuev class includes potentials of
Lenard-Jones type introduced by Fisher (see e.g. \cite{Fisher-Ruelle},
\cite{Rebenko-Tertychnyi} for an overview and
\cite{Lima-Procacci-Yuhjtman} for a proof of this statement).

Basuev \cite{Basuev1} has in addition shown that $e\left( \phi \right) \leq 
\bar{e}\left( \phi \right) \leq 13e\left( \phi \right) /12$ for potentials $
\phi $ such that $\min_{x\in \mathbb{R}^{3}}\phi (\left\vert x\right\vert
)=-\lambda <0$ and has stated that $e\left( \phi \right) =\bar{e}\left( \phi
\right) $ holds for the majority of stable potentials which is useful in
applications. Our result on the equality $e(h)=\bar{e}(h)$ differs, however,
in many respects. Typical Basuev potentials are bounded from below by a
negative constant $-\lambda $, repulsive at short, attractive and integrable
at large distances. Equality in this situation occurs when the infimum in $n$
of (\ref{e}) is attained at $\infty $.\footnote{
Because the minimal specific energy may be written as $\bar{e}=\inf_{n\geq 2}
\frac{n}{n-1}e_{n}$.} The Yukawa potential $v$ on the other hand, as
scale mixtures
of Euclid's hat $h(\left\vert x\right\vert /s)$ weighted by $g(s)$ where $h$,
$g$ and consequently $v$ are all positive functions, repeals (attracts)
two particles with the same (opposite) charges in its entire support. The
infimum in (\ref{e}) is attained for neutral configurations, when $
n_{+}=m\in \mathbb{N}$ positive and $n_{-}=m$ negative charges collapse into
one point while the infimum in the modified specific energy $\bar{e}(v)$ is
attained when $\left\vert n_{+}-n_{-}\right\vert =1$.

\paragraph{One versus iterated Mayer expansion.}

The Ursell functions can be written by the well known formula introduced by
Mayer (see \cite{Uhlenbeck-Ford}): 
\begin{equation}
\psi _{n}^{c}(\zeta _{1},\ldots ,\zeta _{n};v)=\sum_{G~\mathrm{connected}
}\prod_{\langle ij\rangle \in E(G)}\left( \exp \left( -\beta \sigma
_{i}\sigma _{j}v\left( \left\vert x_{i}-x_{j}\right\vert \right) \right)
-1\right) ~,  \label{ursell-functions}
\end{equation}%
where the sum runs over all connected linear graphs $G$ with vertices in the
index set $\left\{ 1,\ldots ,n\right\} $ and $E(G)$ denotes the set of edges
of $G$. As far as the estimation of pressure and correlation functions are
concerned, equation (\ref{ursell-functions}) is not useful due the
cardinality of its sum. To reduce the sum over connected Mayer graphs to
labeled trees, Penrose \cite{Penrose} has exploited cancellations occurring
on the formula under proper re-summation and proved that the Mayer series
converge provided the potential $v$ is stable, integrable at large distances
and has, in addition, a hard core condition which recently has shown
\cite{Procacci-Yuhjtman} to be unnecessary (see also
\cite{Brydges-Martin} for an 
overview and extensions). The cardinality of labeled trees of order $n$ is $
n^{n-2}$ by the famous Cayley theorem, which makes the tree graph identities
suitable for the estimation of thermodynamical functions. Among the proposed
tree graph formulas now available we indicate the one in Theorem 3.1
of \cite{Brydges-Kennedy} as the most adequate to our purposes of
representing the 
Ursell functions $\psi _{n}^{c}(t,\zeta _{1},\ldots ,\zeta _{n})$ defined by
(\ref{ursell-functions}) with the scale--dependent--interaction (\ref{vtx})
in the place of $v$. Such Ursell functions satisfy the system of ordinary
differential equations (\ref{systeqs}).

One particular tree graph identity due to Basuev \cite{Basuev2} is however
worth mentioning in the context of the present work. The
representation of Basuev 
works for radial potentials in $\mathbb{R}^{d}$ of the form $\phi =\phi
^{a}+\delta $ (see definition of Basuev potentials above), where $\phi
^{a}(r)=\phi (r)$ for $r=\left\vert x\right\vert >a$, $\phi ^{a}(r)=\phi (a)$
for $r\leq a$, is stable and $\delta (r)=\phi (r)-\phi ^{a}(r)>0$ for $r\leq
a$ and $\delta (r)=0$ for $r>a$, which may include hard--core: $\delta
(r)=\infty $, $r\leq a$. To estimate the Ursell functions efficiently,
Basuev uses the modified stability condition 
\begin{equation}
U_{n}(\zeta _{1},\ldots ,\zeta _{n};v)\geq -(n-1)\bar{B}  \label{stability1}
\end{equation}%
instead of (\ref{stability}), where in the majority of cases important for
applications $\bar{B}$ is equal or closed to $B$. It might appear that a
slight improvement on the stability bound would not affect the radius of
convergence of Mayer series. It turns out, however, that the estimate of the
Ursell functions through the Basuev tree graph identity works so well
when (\ref{stability1}) is applied (see particularly equations (15)
and (16) of  
\cite{Basuev2}) that expressive improvements on the convergence are reported
(at low temperatures) in Basuev paper, as well as in
\cite{Lima-Procacci-Yuhjtman}. 

Let us now explain how the estimate on the Ursell functions gets improved by
(\ref{stability1}) in our case. It is known that the Mayer series for the
pressure of a two-dimensional Yukawa gas \cite{Brydges-Kennedy, Benfatto,
Guidi-Marchetti} 
\begin{eqnarray}
\beta p(\beta ,z) &=&\sum_{k\geq 1}b_{k}z^{k}  \label{p} \\
b_{k} &=&\frac{1}{k!}\int d^{k-1}\varrho \psi _{k}^{c}(\zeta _{1},\ldots
,\zeta _{k};v)  \notag
\end{eqnarray}%
converges if $\left\vert z\right\vert <(4\pi -\beta )/(4\pi e\beta )$, the
radius of convergence being positive provided $\beta <4\pi $. Since the
Yukawa potential (\ref{v}) diverges logarithmically as $\left\vert
x\right\vert \rightarrow 0$, the proof of such statement requires the use of
Brydges--Kennedy approach or iterated Mayer expansion (no one-scale tree
expansion formula would be able to deal with this issue). The problem at our
hand is to extend the stability of $2$--dimensional Yukawa gas to the
inverse temperature in the range $4\pi \leq \beta <8\pi $, passing through
the sequence of thresholds $\beta _{2r}=8\pi \left( 1-1/2r\right) $, $r\in 
\mathbb{N}$. Here, $\beta _{2r}$ is the inverse temperature in which a
clusters with $r$ positive and $r$ negative charges collapse altogether at
once, heuristically given by an argument of entropy--energy (there are $
r^{2} $ and $r(r-1)$ distinct pairings of opposite, respectively, same
charges):
\begin{equation}
C(\delta )=\int_{\left\vert x_{2}\right\vert \leq \delta }dx_{2}\cdots
\int_{\left\vert x_{2r}\right\vert \leq \delta }dx_{2r}\exp \left( \beta
\left( r^{2}-r(r-1)\right) \int_{\delta }^{1}g(s)ds\right) ~.  \label{delta}
\end{equation}%
By the second mean value theorem and $g(s)\asymp 1/(2\pi s)$ as $
s\rightarrow 0$, the balance expressed by (\ref{delta}) is in favor of
entropy $S(\delta )=\delta ^{2(2r-1)}$ if $\beta <\beta _{2r}$ while the
energy contribution $e^{-\beta E(\delta )}\simeq \delta ^{-\beta r/(2\pi )}$
dominates if $\beta >\beta _{2r}$ so we have%
\begin{equation*}
\lim_{\delta \rightarrow 0}C(\delta )=c\lim_{\delta \rightarrow 0}\delta
^{2(2r-1)-\beta r/(2\pi )}=\left\{ 
\begin{array}{cc}
0 & \text{if }\beta <\beta _{2r} \\ 
\infty & \text{if }\beta >\beta _{2r}%
\end{array}%
\right. ~,
\end{equation*}%
for some constant $c>0$.

\paragraph{Avoiding the collapse of neutral clusters: a conjecture}

As a consequence of the alluded collapses, the leading even coefficients $
b_{2j}$, $j=1,\ldots ,n$, of the Mayer series (\ref{p}) diverges for $\beta
_{2n}\leq \beta <\beta _{2(n+1)}$ when the short scale cutoff $t_{0}$,
introduced in (\ref{vtx}) (or in (\ref{v1})) to make the system
conditionally stable, is removed. A conjecture stated as an open problem in 
\cite{Benfatto} is as follows:

\begin{conjecture}
\label{conjecture}If the leading $n$ even coefficients $b_{2j}$'s are
removed from the Mayer series (\ref{p}), the radius of convergence of the
corresponding series remains positive for any $\beta \in \lbrack \beta
_{2n},\beta _{2(n+1)})$ and, consequently, for any $\beta <\beta _{2(n+1)}$.
\end{conjecture}

Brydges-Kennedy \cite{Brydges-Kennedy} have proved convergence of (\ref{p})
with $O\left( z^{2}\right) $ term omitted for $4\pi \leq \beta <16\pi /3$
and have explained how it would be extended up to the second threshold $6\pi 
$. It turns out that the claimed improvement on the estimate of the
three--particle energy from $U_{3}(\xi _{1},\xi _{2},\xi _{3};\dot{v})\geq -3%
\dot{v}(t,0)/2$ to $U_{3}(\xi _{1},\xi _{2},\xi _{3};\dot{v})\geq -\dot{v}%
(t,0)$ does not hold uniformly on $\left( \left\{ -1,1\right\} \times 
\mathbb{R}^{2}\right) ^{3}$ at each scale for the decomposition (\ref{v1})
used by the authors. It has been shown by numerical calculation in
\cite{Guidi-Marchetti} that the factor $3$ (the number $n$ of charged
particles 
involved) in the lower bound of $U_{3}$ may be improved to $2.14..$., which
is enough to extend the convergence of Mayer series to any $\beta \in
\lbrack 4\pi ,6\pi )$ but insufficient to establish the conjecture beyond a
certain threshold (about $\beta _{15}=112\pi /15$) up to $8\pi $. For the
latter, it is indeed necessary to improve the factor from $3$ to $2$ ($n=3$
to $n-1=2$). Both statements are proved in the present work. In addition, we
have proved that, if the decomposition (\ref{vgh}) for the Yukawa potential
is used instead, then by (\ref{UnSS}) $3$ can be replaced by $2$ in the
stability bound for $U_{3}$ and for every odd $n$ (\ref{stability}) can be
substituted by (\ref{stability1}) with $B=\bar{B}=v(0)/2$.

The purpose of the present paper is also to provide a majorant candidate for
the pressure of the Yukawa gas at $\beta _{2n}\leq \beta <\beta _{2(n+1)}$,
uniformly in the cutoff $t_{0}$, when it is extracted from the even leading
Mayer coefficients $b_{2j}$, $j=1,\ldots ,n$, their divergent part. Such
majorant has been proposed in \cite{Guidi-Marchetti} but our presentation is
neater than the original paper making it more transparent. We adapt to the
potential decomposition (\ref{vgh}) all ingredients and the construction
used in that reference through the scale decomposition (\ref{v1}).

The majorant construction is based on the idea already present in the early
works by Imbrie \cite{Imbrie} and \cite{Brydges}, according to which the
Mayer series (\ref{p}) (after some combinatorics together with the stability
estimate) is dominated by an expansion in powers of $ez\left\Vert \beta
v\right\Vert _{1}e^{B}$, where $\left\Vert \beta v\right\Vert _{1}$ is the $
L^{1}$--norm of $\beta v(x)$ and $B=\beta v(0)/2$. A one--step Mayer
expansion is not suitable to potentials that $B$ is large in the range that $
\left\Vert \beta v\right\Vert _{1}$ contributes little, as typically occurs
for the two--dimensional Yukawa potential $v$ (see \cite{Imbrie,
Gopfert-Mack, Brydges-Kennedy} for other applications). When $v$ is
decomposed into a continuum of scales (see (\ref{v1}), or
alternatively (\ref{vtx})), the Mayer series becomes, roughly
speaking, an expansion in powers of $ez\tau (t_{0},t)$, where 
\begin{equation}
\tau (t_{0},t)=\displaystyle\int_{t_{0}}^{t}\left\Vert \beta \dot{v}(s,\cdot
)\right\Vert _{1}e^{\beta \int_{s}^{t}\dot{v}(\tau ,0)d\tau }ds~
\label{tau-1}
\end{equation}
solves a linear equation (\ref{C2}) satisfied by the majorant $C_{2}$ of two
times the second Mayer coefficient: $2\left\vert b_{2}\right\vert \leq C_{2}$
(see (\ref{bnAn}) and (\ref{nAnCn})). It has been shown that the Mayer
expansion (\ref{p}) converges provided $\beta \in \lbrack 0,4\pi )$ and $
e\left\vert z\right\vert \tau (t_{0},t)<1$ uniformly in $t_{0}>0$ (see
Theorem 4.1 together with pgs. 41-42 of \cite{Brydges-Kennedy} and
Proposition \ref{classical}, Remarks \ref{4pi} and \ref{r4pi} below). Inside
the first threshold, the domain of convergence is replaced by $\left( \beta
,z\right) \in \mathbb{R}_{+}\times \mathbb{C}$ such that  $\beta \in $ $
[4\pi ,16\pi /3)$ and $e\left\vert z\right\vert \displaystyle
\int_{t_{0}}^{t}\left\Vert \beta \dot{v}(s,\cdot )\right\Vert _{1}e^{(3\beta
/2)\int_{s}^{t}\dot{v}(\tau ,0)d\tau }ds<1$ and we shall see that our
candidate to majorant series converges provided $\beta \in \lbrack \beta
_{k},\beta _{k+1})$ and the factor $3\beta /2$ in the exponent of this
domain is replaced by $(k+1)\beta /k$ for any $k>1$. If $C_{n}$, $1<n\leq k$
, denote the first $k-1$ majorant coefficients: $n\left\vert
b_{n}\right\vert \leq C_{n}$, we observe by (\ref{Cn1k}) that $(k+1)(n-1)B/k$
multiplies the linear term of the equation satisfied by $C_{n}$ (after the
divergent part of the even $n\leq k$ coefficients have been extracted
through a Lagrage multiplier $L_{k}$). In particular, for $n=2$, $\tau
_{k}(t_{0},t)$ given by (\ref{tauk}) generalizes (\ref{tau-1}) and solves
the linear equation (\ref{C2k}) for $C_{2}$. Since the modified stability
condition (\ref{stability1}) applies for every $n>1$ odd, the coefficient
that multiplies the linear term of the equation for $C_{n}$, which is given
by $(n-1)B<(k+1)(n-1)B/k$, implies that the same equation satisfied by $C_{n}
$ with $n$ even holds for $n$ odd. To understand why the modified stability
bound (\ref{stability1}) is so crucial, we observe that anything large than $
(n-1)B$ would prevent the convergence of the majorant series in the whole
interval of collapse $[4\pi ,8\pi )$. Recall that, when the standard scale
decomposition is used, the interacting energy $U_{n}$ with $n=3$ is bounded
below by a factor $-2.14B$, instead of $-2B$, preventing the $nb_{n}$ to be
dominated by $C_{n}$ for $\beta >\beta _{k}$ with $k$ verifying the
inequality $2.14>2(k+1)/k$, i. e. $k>15$.

\paragraph{Outlines of the present work}

The present paper is organized as follows. Section \ref{MSE} is dedicated to
the proof of the main Theorem \ref{main} and Corollary \ref{minimal}
together with estimates on the modified Bessel functions of second kind
envolved in both representations of two-dimensional Yukawa potential:
standard (Proposition \ref{bare3}) and scale mixtures of Euclid's hat
(Proposition \ref{exponential}).

The main Theorem is then used to resolve an obstructive remark of an
umpublished paper (Remark 7.5 of \cite{Guidi-Marchetti}) which, whether the
standard decomposition of the Yukawa potential into scales were adopted,
would impede a direct proof of the convergence of the Mayer series of the
two-dimensional Yukawa gas for the inverse temperature up to $8\pi $. We
dedicate Section \ref{MDF} to the Cauchy majorant method applied to the
density function of Yukawa gas on the whole interval $[4\pi ,8\pi )$ of
collapses. For this system it is proven that the Mayer series converge up to
the second threshold $\beta \in \lbrack 4\pi ,6\pi )$ and its given
explanations on the mechanism that allows it to be extended up to $8\pi $.
The paper distinguishes the stability issues from those matters related to
convergence of the Mayer series. In respect to the former its is proven at
the end of Section \ref{MDF} that dipoles in the presence of other charges
are prevented to collapse.

\section{Minimal specific energy: main theorem and estimates involving
modified Bessel functions \label{MSE}}

\setcounter{equation}{0} \setcounter{theorem}{0}

We prove in this section our main theorem (\ref{UnSS}) and the implications
of it on the minimal specific energies $e(h)$ and $\bar{e}(h)$ for the
Euclid's hat $h$ in $\mathbb{R}^{2}$.

\paragraph{Three particles minimal specific energy}

To begin with, let $U_{n}$ be the $n$--particle total energy (\ref{Unzetas})
and let 
\begin{equation}
e_{n}(v)=\frac{1}{n}\inf_{\substack{ \left( \zeta _{1},\ldots ,\zeta
_{n}\right) ,  \\ \zeta _{i}=(\sigma _{i},x_{i})\in \left\{ -1,1\right\}
\times \mathbb{R}^{2}}}U_{n}(\zeta _{1},\ldots ,\zeta _{n};v)  \label{env}
\end{equation}%
and%
\begin{equation}
\bar{e}_{n}(v)=\frac{1}{n-1}\inf_{\substack{ \left( \zeta _{1},\ldots ,\zeta
_{n}\right) ,\zeta _{i}\in \left\{ -1,1\right\} \times \mathbb{R}^{2}:  \\ %
\sigma _{1}+\cdots +\sigma _{n}\neq 0}}U_{n}(\zeta _{1},\ldots ,\zeta _{n};v)
\label{renv}
\end{equation}%
be the $n$--particles minimal, and constrained minimal, specific energies.
As the particles of our system have either $+1$ or $-1$ charges, these two
quantities are related to each other when $n$ is an odd number as $e_{n}(v)=%
\bar{e}_{n}(v)(n-1)/n$. Let us first consider the case $n=3$ and let $v(x)$
be given by the two--dimensional Yukawa potential (\ref{v}) under the
standard decomposition into scales (\ref{v1}), cut-off at short distances $%
s\leq t_{0}$: $v(x)=\displaystyle\int_{t_{0}}^{1}\tilde{h}(\left\vert
x\right\vert /s)/(2\pi s)~ds$. Since, by (\ref{ev}),%
\begin{equation*}
\bar{e}_{n}(v)=\int_{t_{0}}^{1}\frac{1}{2\pi s}~\bar{e}_{n}\left( \tilde{h}%
(\cdot /s)\right) ~ds=\frac{1}{2\pi }\log \frac{1}{t_{0}}~\bar{e}_{n}\left( 
\tilde{h}\right) ~,
\end{equation*}%
it is enough to consider the minimal specific energy of $3$--particles $\bar{%
e}_{3}(\tilde{h})$ for $\tilde{h}(w)=wK_{1}(w)$, where $K_{1}$ is the
modified Bessel function of second kind of order $1$.

We shall need among other properties some general features of $\tilde{h}(w)$.

\begin{proposition}
\label{h1}$w\longmapsto \tilde{h}(w)=wK_{1}(w)$ is a regular function at
every point $w\in \left( 0,\infty \right) $. The function $\tilde{h}(w)$
strictly decreases from its maximum $\tilde{h}(0)=1$, decays to $0$ at $%
\infty $ exponentially fast and changes its concavity: $\tilde{h}^{\prime
\prime }(w)<0$ for $w<w_{0}$ and $\tilde{h}^{\prime \prime }(w)>0$ for $%
w>w_{0}$ at $1/2<w_{0}<\left( 1+\sqrt{17}\right) /8$ whose numerical value
is $w_{0}=0.5950$(\ldots ).
\end{proposition}

\begin{figure}[th]
\centering
\includegraphics[scale=0.45]{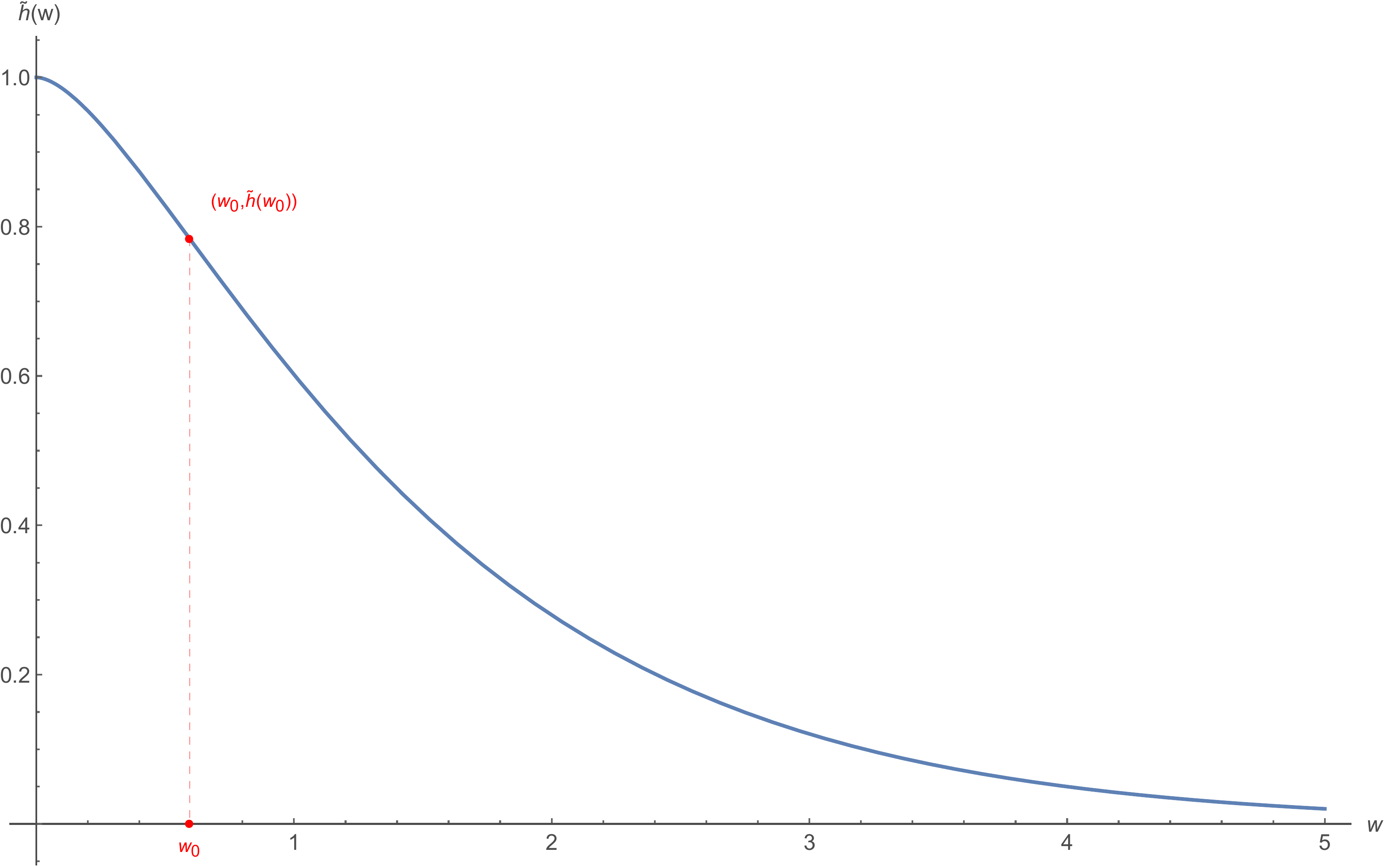} 
\centering
\caption{Plot of $\tilde{h}(w)$. }
\label{fig1}
\end{figure}

\noindent \textit{Proof.} Regularity and positivity of $K_{\nu }(x)$ for
every $\nu \in \mathbb{R}$ and $x>0$ are known facts (see e.g. Appendix A of 
\cite{Gaunt}). It follows from the equation
\begin{equation}
\left( x^{n}K_{n}(x)\right) ^{\prime }=-x^{n}K_{n-1}(x)  \label{xnKn}
\end{equation}%
with $n=1$ together with $\lim_{w\rightarrow 0}wK_{0}(w)=0$ and $%
\lim_{w\rightarrow 0}wK_{1}(w)=1$ (see \cite{Gaunt} and Lemma 2.2 of \cite%
{Yang-Chu}) that%
\begin{equation*}
\tilde{h}^{\prime }(w)=\left( wK_{1}(w)\right) ^{\prime }=-wK_{0}(w)<0
\end{equation*}%
for $w>0$, proving the strictly decreasing property of $\tilde{h}$ and $%
\tilde{h}(0)=1$. The inequalities for $x>0$: 
\begin{eqnarray}
\frac{\sqrt{\pi }e^{-x}}{\sqrt{2x+1/2}} &<&K_{0}(x)<\frac{\sqrt{\pi }e^{-x}}{%
\sqrt{2x}}  \notag \\
1+\frac{1}{2x+1/2} &<&\frac{K_{1}(x)}{K_{0}(x)}<1+\frac{1}{2x}\ ,  \label{KK}
\end{eqnarray}%
find in ref. \cite{Yang-Chu}, imply the exponential decaying of $\tilde{h}%
(w) $ and together with%
\begin{eqnarray*}
\tilde{h}^{\prime \prime }(w) &=&-\left( wK_{0}(w)\right) ^{\prime } \\
&=&wK_{1}(w)-K_{0}(w) \\
&=&K_{0}(w)\left( \frac{wK_{1}(w)}{K_{0}(w)}-1\right)
\end{eqnarray*}
(\ref{KK}) yield%
\begin{equation*}
\tilde{h}^{\prime \prime }(w)<K_{0}(w)\left( w-\frac{1}{2}\right) <0
\end{equation*}%
provided $w<1/2$ and%
\begin{equation*}
\tilde{h}^{\prime \prime }(w)>K_{0}(w)\left( \frac{w}{2w+1/2}+w-1\right) >0
\end{equation*}%
provided $w>(2w+1/2)(1-w)=3w/2+1/2-2w^{2}$ or, equivalently, $w>\left( 1+%
\sqrt{17}\right) /8=0.640\,39$. The unique solution of $%
wK_{1}(w)/K_{0}(w)-1=0$, whose numerical value is $w_{0}=0.5950$(\ldots ),
satisfies $1/2<w_{0}<0.640\,39$ (see proof of Lemma \ref{px}). This
concludes the proof.

\hfill $\Box $

Because the particles interact via a pair potential, it is easy to see that
the minimum potential energy is due to a system in which two of the three
particles have equal signs and the third has charge with the opposite sign.
The potential energy (\ref{Unzetas}) with $n=3$ and $\sigma _{1}=\sigma
_{3}=-\sigma _{2}$ is then given by%
\begin{equation*}
U_{3}(\zeta _{1},\zeta _{2},\zeta _{3};\tilde{h})=-\tilde{h}\left(
\left\vert x_{1}-x_{2}\right\vert \right) -\tilde{h}\left( \left\vert
x_{2}-x_{3}\right\vert \right) +\tilde{h}\left( \left\vert
x_{1}-x_{3}\right\vert \right) ~.
\end{equation*}%
To simplify the expression, we write $r_{1}=\left\vert
x_{1}-x_{2}\right\vert $, $r_{2}=\left\vert x_{2}-x_{3}\right\vert $ and $%
r_{3}=\left\vert x_{1}-x_{3}\right\vert $ can be written, as the particles
are located at the vertices of a triangle, by the law of cosine, as%
\begin{equation*}
r_{3}(r_{1},r_{2},\theta )=\sqrt{\left( r_{1}-r_{2}\right)
^{2}+4r_{1}r_{2}\sin ^{2}\theta /2}~.
\end{equation*}%
Since $\tilde{h}(w)$ is a strictly decreasing function, the minimal specific
energy of $3$--particles (\ref{env}) thus reads 
\begin{eqnarray}
\bar{e}_{3}(\tilde{h}) &=&\frac{1}{2}\min_{r_{1},r_{2}\geq 0,0\leq \theta
\leq \pi }\left( \tilde{h}(r_{3}(r_{1},r_{2},\theta ))-\tilde{h}(r_{1})-%
\tilde{h}(r_{2})\right)  \notag \\
&=&\frac{1}{2}\min_{r_{1},r_{2}\geq 0}\left( \tilde{h}(r_{1}+r_{2})-\tilde{h}%
(r_{1})-\tilde{h}(r_{2})\right) ~.  \label{ebar3}
\end{eqnarray}%
The next proposition shows that this quantity does not reach from below the
value $-1/2=\left( -\sum_{i=1}^{3}\sigma _{i}^{2}+\left\vert
\sum_{i=1}^{3}\sigma _{i}\right\vert \right) /(2\cdot (3-1))$ that one would
expected for a convex function $h$.

\begin{proposition}
\label{bare3}%
\begin{equation}
\left( K_{1}(1)-K_{1}(1/2)\right) /2>\bar{e}_{3}(\tilde{h})>-0.535\ .
\label{e3}
\end{equation}
\end{proposition}

\begin{remark}
\label{e3estimate}As the numerical evaluations used in the proofs are sharp
up to high decimal order, we may claim that $\bar{e}_{3}=-0.530$(\ldots ),
which is certainly less than $-1/2$ ($-0.527$(...)$>\bar{e}_{3}>-0.535$),
according to the precision of the machine used to calculate it.
\end{remark}

\noindent \textit{Proof.} To prove (\ref{e3}), it is enough by (\ref{ebar3})
to show that%
\begin{equation}
\tilde{h}(x+y)-\tilde{h}(x)-\tilde{h}(y)+1.07>0  \label{hhh}
\end{equation}%
holds for all $x$, $y\geq 0$. Defining $f(x)=\tilde{h}(x)-1.07$, equation (%
\ref{hhh}) is equivalent to show superadditivity of $f(x)$:%
\begin{equation}
f(x+y)>f(x)+f(y)~.  \label{fff}
\end{equation}%
But this is implied by the following

\begin{lemma}
\label{super}Let $q(x)=f(x)/x$ be defined for $x>0$. If $q(x)$ is monotone
increasing, then $f(x)$ is superadditive.
\end{lemma}

\noindent \textit{Proof of Lemma \ref{super}.} Suppose that $g(x)$ is
monotone increasing function. Then $q(x+y)\geq q(x)$, $q(x+y)>q(y)$ and it
follows that%
\begin{eqnarray*}
f(x+y) &=&x\frac{f(x+y)}{x+y}+y\frac{f(x+y)}{x+y} \\
&=&xq(x+y)+yq(x+y) \\
&>&xq(x)+yq(y) \\
&=&f(x)+f(y)~,
\end{eqnarray*}%
which proves the lemma.

\hfill $\Box $

It remains thus to prove that $q(x)=(\tilde{h}(x)-1.07)/x=K_{1}(x)-1.07/x$
is monotone increasing. From (\ref{xnKn}) with $n=1$, we deduce%
\begin{equation*}
K_{1}(x)+xK_{1}^{\prime }(x)=\left( xK_{1}(x)\right) ^{\prime }=-xK_{0}(x)
\end{equation*}%
which implies that%
\begin{eqnarray*}
q^{\prime }(x) &=&K_{1}^{\prime }(x)+\frac{1.07}{x^{2}} \\
&=&\frac{-1}{x^{2}}\left( xK_{1}(x)+x^{2}K_{0}(x)-1.07\right) >0
\end{eqnarray*}%
for $x>0$ provided%
\begin{equation*}
xK_{1}(x)+x^{2}K_{0}(x)<1.07~.
\end{equation*}%
This inequality, however, holds in view of the following:

\begin{lemma}
\label{px}The function $x\longmapsto p(x)=xK_{1}(x)+x^{2}K_{0}(x)$ defined
in $\mathbb{R}_{+}$ has a global maximum at $x_{0}$, $1/2<x_{0}<\left( 1+
\sqrt{17}\right) /8$. It strictly increases with $p(0)=1$ as $x$ varies from 
$0$ to $1/2$ and strictly decreases to $0$, exponentially fast, as $x$
varies from $\left( 1+\sqrt{17}\right) /8$ to $\infty $. The second
derivative $p^{\prime \prime }(x)$ of $p(x)$ is negative in the interval $
1/2\leq x_{0}\leq \left( 1+\sqrt{17}\right) /8$. Numerically, $x_{0}=0.5950$
(\ldots ) and its (global) maximum values $p(x_{0})=1.061$(\ldots )$~<1.07$.
\end{lemma}

\begin{figure}[th]
\centering
\includegraphics[scale=0.45]{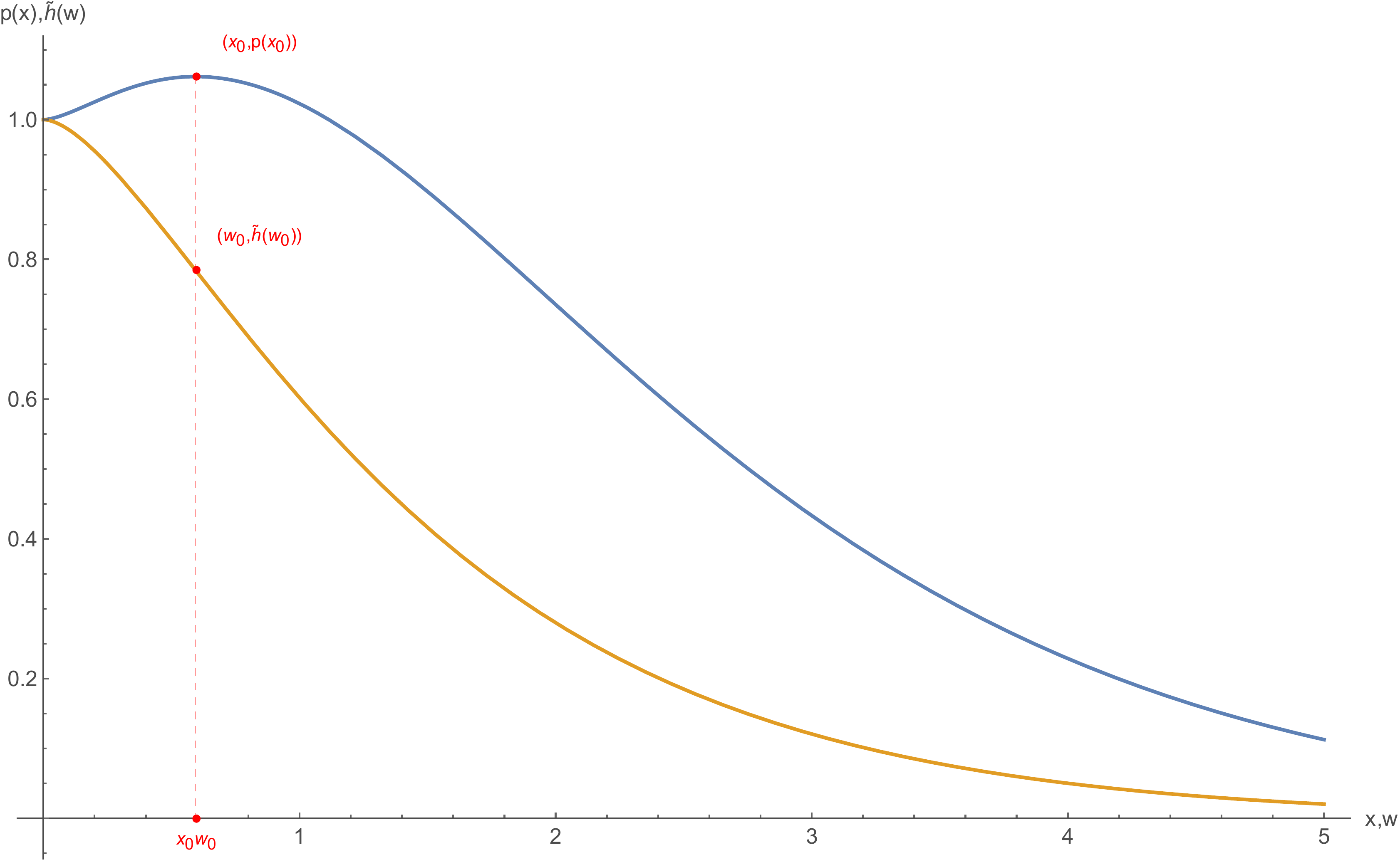} 
\centering
\caption{Plot of $p(x)$ and $\tilde{h}(w)$ together. }
\label{fig2}
\end{figure}

\noindent \textit{Proof of Lemma \ref{px}.} Using (\ref{xnKn}) with $n=1$
together with $K_{0}^{\prime }(x)=-K_{1}(x)$, as in the proof of Proposition %
\ref{h1}, we have%
\begin{eqnarray*}
p^{\prime }(x) &=&-xK_{0}(x)+2xK_{0}(x)-x^{2}K_{1}(x) \\
&=&xK_{0}(x)\left( 1-\frac{xK_{1}(x)}{K_{0}(x)}\right) \\
&<&xK_{0}(x)\left( 1-\frac{x}{2x+1/2}-x\right) <0
\end{eqnarray*}%
provided $x>\left( 1+\sqrt{17}\right) /8$ and 
\begin{equation*}
p^{\prime }(x)>xK_{0}(x)\left( x-\frac{1}{2}\right) >0
\end{equation*}%
provided $x<1/2$. These prove that $p(x)$ increases in $\left( 0,1/2\right) $
and decreases in $\left( \left( 1+\sqrt{17}\right) /8,\infty \right) $,
exponentially fast in view of (\ref{KK}).

$p(x)$ attains its maximum value at the same point at which $\tilde{h}(w)$
changes its concavity. The maximum $x_{0}$ of $p(x)$ solves $%
K_{0}(x)-xK_{1}(x)=0$ and satisfies $1/2<x_{0}<$ $\left( 1+\sqrt{17}\right)
/8\approx \allowbreak 0.64$, as stated above and showed in Proposition \ref%
{h1}. To prove that $x_{0}$ is the global maximum, it suffices to show that
the second derivative of $p(x)$, which may be calculated exactly as before,%
\begin{eqnarray*}
p^{\prime \prime }(x) &=&\left( x\left( K_{0}(x)-xK_{1}(x)\right) \right)
^{\prime } \\
&=&K_{0}(x)-xK_{1}(x)+x(K_{0}^{\prime }(x)-\left( xK_{1}(x)\right) ^{\prime
}) \\
&=&(1+x^{2})K_{0}(x)-2xK_{1}(x) \\
&=&-2K_{0}(x)\left( \frac{xK_{1}(x)}{K_{0}(x)}-\frac{1+x^{2}}{2}\right) \ ,
\end{eqnarray*}%
takes negative values for $x\in \left[ 1/2,\left( 1+\sqrt{17}\right) /8%
\right] $. By equation (\ref{KK}) and positivity of $K_{0}(x)$, this is
implied by 
\begin{equation*}
\frac{xK_{1}(x)}{K_{0}(x)}-\frac{1+x^{2}}{2}>x+\frac{x}{2x+1/2}-\frac{1+x^{2}%
}{2}>0~.
\end{equation*}%
Denoting the function on the right hand side by $%
l(x)=x+x/(2x+1/2)-(1+x^{2})/2$, we need to show that $l(x)>0$ for $x\in %
\left[ 1/2,\left( 1+\sqrt{17}\right) /8\right] $. But $l(1/2)=5/24\approx
0.20$ and $l(\left( 1+\sqrt{17}\right) /8)=\left( 23+\sqrt{17}\right)
/64\approx 0.29$ are both positive and the second derivative of $l(x)$, 
\begin{equation*}
l^{\prime \prime }(x)=-\left( 17+12x+48x^{2}+64x^{2}\right) /(1+4x)^{3}<0
\end{equation*}%
for all $x>0$, proving therefore the statement.

We have thus proven that $x_{0}$ is a global maximum of $p(x)$, concluding
the proof of Lemma \ref{px}.

\hfill $\Box $

Returning to the proof of Proposition \ref{bare3}, the lower bound stated in
(\ref{e3}) follow from the superadditivity of $f(x)=\tilde{h}(x)-1.07$,
which is proven in Lemmas \ref{super} and \ref{px}. The numerical estimate
for the specific energy $\bar{e}_{3}$ stated in Remark \ref{e3estimate} is
obtained when $1.07$ is replaced by the maximum values $p(x_{0})=1.061$%
(\ldots ), given in Lemma \ref{px}, since at this point $f(x_{0})=\tilde{h}%
(x_{0})-1.061$(\ldots ) satisfies (\ref{fff}) as an equality and
consequently, by (\ref{hhh}), $\tilde{h}(2x_{0})-\tilde{h}(x_{0})-\tilde{h}%
(x_{0})=-1.061(\ldots )$.

By definition (\ref{ebar3}), taking $r_{1}=r_{2}=1/2$ in the expression
inside the minimum, we have an upper bound
\begin{equation*}
\bar{e}_{3}<\frac{1}{2}\left( \tilde{h}(1)-2\tilde{h}(1/2)\right) =\frac{1}{2%
}(K_{1}(1)-K_{1}(1/2))=-0.527(...)~.
\end{equation*}

\hfill $\Box $

\begin{remark}
\label{mayer-e3}It does not seem easy to extend the superadditivity method
used to estimated the (restricted) minimum specific energy of $3$--particles
to $(2k+1)$--particles with $k>1$. As we shall see in the next section, the
result on the minimal specific energy $\bar{e}_{3}\left( \tilde{h}\right) $
prevents that the third Mayer coefficient be defined uniformly in the cutoff 
$t_{0}$ in the entire collapse interval $\left[ 4\pi ,8\pi \right] $,
although it is enough for concluding convergence of the Mayer series up to
the second threshold $[4\pi ,6\pi )$. Numerical calculations performed in 
\cite{Guidi-Marchetti} indicate that $\bar{e}_{2k+1}\left( \tilde{h}\right) $
remains for $k>1$ strictly smaller than $-1/2$. We should mention that if $%
\tilde{h}(w)$ were convex, the minimal of (\ref{ebar3}) would be attained at 
$r_{1}=r_{2}=0$, obtaining the expected value $\bar{e}_{3}=-1/2$ as it is
exactly the case when we use decomposition (\ref{vgh}) of the Yukawa
potential (\ref{v}) instead of (\ref{v1}). Since the method based on
superadditivity cannot be easily extended to $k>1$, another method will be
employed to obtain $\bar{e}_{2k+1}(h)=-1/2$ with $h$ the Euclid's hat
function (\ref{h}).
\end{remark}

\paragraph{The main theorem}

We shall now turn to the representation of Yukawa potential (\ref{v}) given
by $v(x)=v_{(0,\infty )}(x)=K_{0}(\left\vert x\right\vert )/(2\pi )$ where
(see (\ref{vgh})):%
\begin{equation}
v_{\left( t_{0},t\right) }(x)=\int_{t_{0}}^{t}h(\left\vert x\right\vert
/s)g(s)ds~,  \label{vt0t}
\end{equation}%
is a scale mixtures of Euclid's hat. Here, for $x\in \mathbb{R}^{2}$ and $%
s\in \mathbb{R}_{+}$,%
\begin{equation}
h(\left\vert x\right\vert /s)=\frac{4}{\pi s^{2}}\chi _{\lbrack 0,s/2]}\ast
\chi _{\lbrack 0,s/2]}(x)  \label{hs}
\end{equation}%
is the self convolution of indicator function $\chi _{\lbrack
0,s/2]}(x):=\theta (s/2-\left\vert x\right\vert )$ of the $2$--dimensional
ball (disc) $B_{r}\equiv B_{r}(0)$ of radius $r=s/2$ centered at origin and $%
g(s)$ is the scale mixtures density given by Hainzl--Seiringer: \cite%
{Hainzl-Seiringer}%
\begin{equation}
g(s)=\frac{-s}{4\pi }\int_{s}^{\infty }K_{0}^{\prime \prime \prime }(r)\frac{%
r}{\sqrt{r^{2}-s^{2}}}dr~.  \label{gs}
\end{equation}

\begin{figure}[th]
\centering
\includegraphics[scale=0.45]{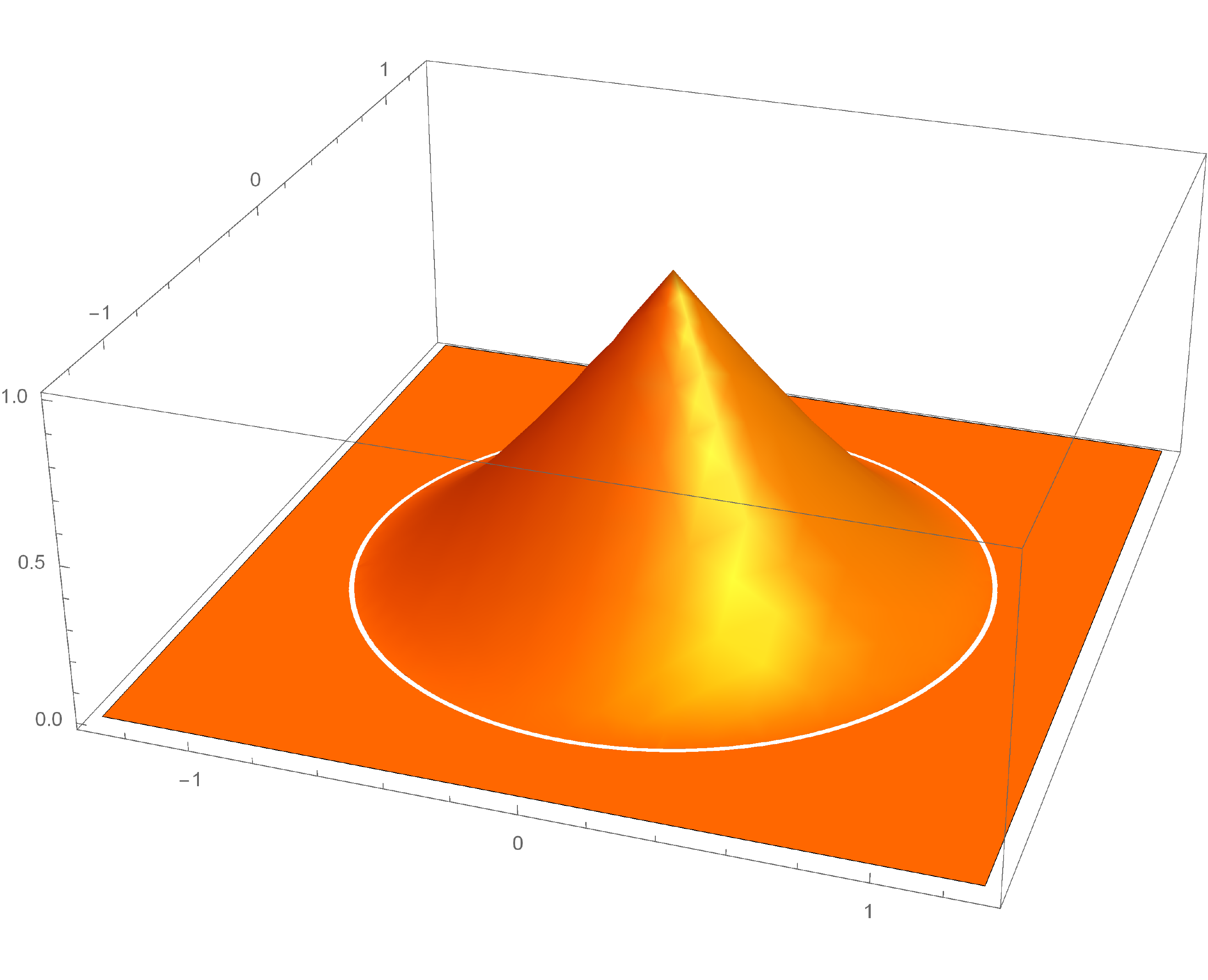} 
\centering
\caption{Euclid's hat function. }
\label{fig3}
\end{figure}

We observe that (\ref{gs}) differs from the $g(s)$ in equation (11) of
\cite{Hainzl-Seiringer} by a pre--factor $\pi \left( s/2\right) ^{2}$
that we 
have used in (\ref{hs}) in order to normalize $h$ at origin: $h(0)=1$. This
normalization is suitable when the radial function $\varphi (\left\vert
x\right\vert )=v(x)$ is the characteristic function of a spherically
symmetric probability distribution in $\mathbb{R}^{d}$ or the covariance of
a stationary and isotropic random field on $d$--dimensional Euclidean space.
The latter is the point of view of the present paper, while the former were
the focus of Gneiting paper \cite{Gneiting}, for which the classes $H_{d}$
of radial positive definite functions generated by scale mixtures of $d$%
--dimensional Euclid's hat $h_{d}(\left\vert x\right\vert )$ played an
important role in the proof of an analogue of Pólya's criterion for $d>1$.
We observe however that the scale mixture used in \cite{Gneiting} is of the
form $\varphi (t)=\displaystyle\int_{0}^{\infty }h_{d}(rt)dG(r)$, where $%
G(r) $ is a probability distribution function in $\left( 0,\infty \right) $
with $G(0+)=c\in \left[ 0,1\right] $. In order to compare with our $g(s)$
given by (\ref{gs}) (by (\ref{vt0t}) $s=1/r$), which behaves as $s$ goes to $%
0$ as $1/(2\pi s)$, in the case that $dG(r)$ is absolutely continuous and
finite positive measure in $\mathbb{R}_{+}$, we write $%
dG(r)=f(r)dr=-f(1/s)ds/s^{2}=-\tilde{f}(s)ds$. We see that $\tilde{f}(s)\sim
1/s$ would lead to a nonintegrable mixture density $f(r)\sim 1/r$ at
infinity and, consequently, $\varphi (t)$ with such a density would not
belong to the class $H_{2}$ considered in that paper.

Equation (\ref{gs}) can be written in terms of a Meijer $G$--functions that
is regular at $s=0$ as 
\begin{equation}
2\pi sg(s)=\sqrt{\pi }G_{13}^{30}\left( s^{2} /4 \left\vert \genfrac{}{}{0pt}{1}{
1/2}{0,1,2}\right. \!\right) ~  \label{g-s}
\end{equation}%
as one can check using Mathematica program together with the shift property: 
$t^{2}G_{13}^{30}\left( t\left\vert
    \genfrac{}{}{0pt}{1}{-3/2}{-2,-1,0}\right. \!\right) 
=G_{13}^{30}\left( t\left\vert
    \genfrac{}{}{0pt}{1}{1/2}{0,1,2}\right. \!\right) $. 

We begin by describing the general features of $h(w)$. We shall state and
prove our main theorem afterwards and return to the asymptotic properties of
(\ref{g-s}) required for the next section.

\begin{proposition}
\label{euclidhat}$w\longmapsto h(w)$ defined by (\ref{hs}) is regular at
every point $w\in \left( 0,1\right) $, convex and non increasing function in 
$\left( 0,\infty \right) $. Moreover, it can be written as
\begin{equation}
h(w)=\frac{2}{\pi }\left( \arccos w-w\sqrt{1-w^{2}}\right) ~,\qquad \text{if}
\ \quad 0\leq w\leq 1  \label{hw}
\end{equation}%
$h(w)=0$ if $w>1$ so, writing $\varphi (x)=h(\left\vert x\right\vert )$ we
have $\varphi (0)=h(0)=1$ and $\hat{\varphi}(0)=\displaystyle\int_{\mathbb{R}%
^{2}}h(\left\vert x\right\vert )dx=\pi /4$ is its Fourier transform $\hat{%
\varphi}(\xi )$ at $\xi =0$.
\end{proposition}

\begin{figure}[th]
\centering
\includegraphics[scale=0.45]{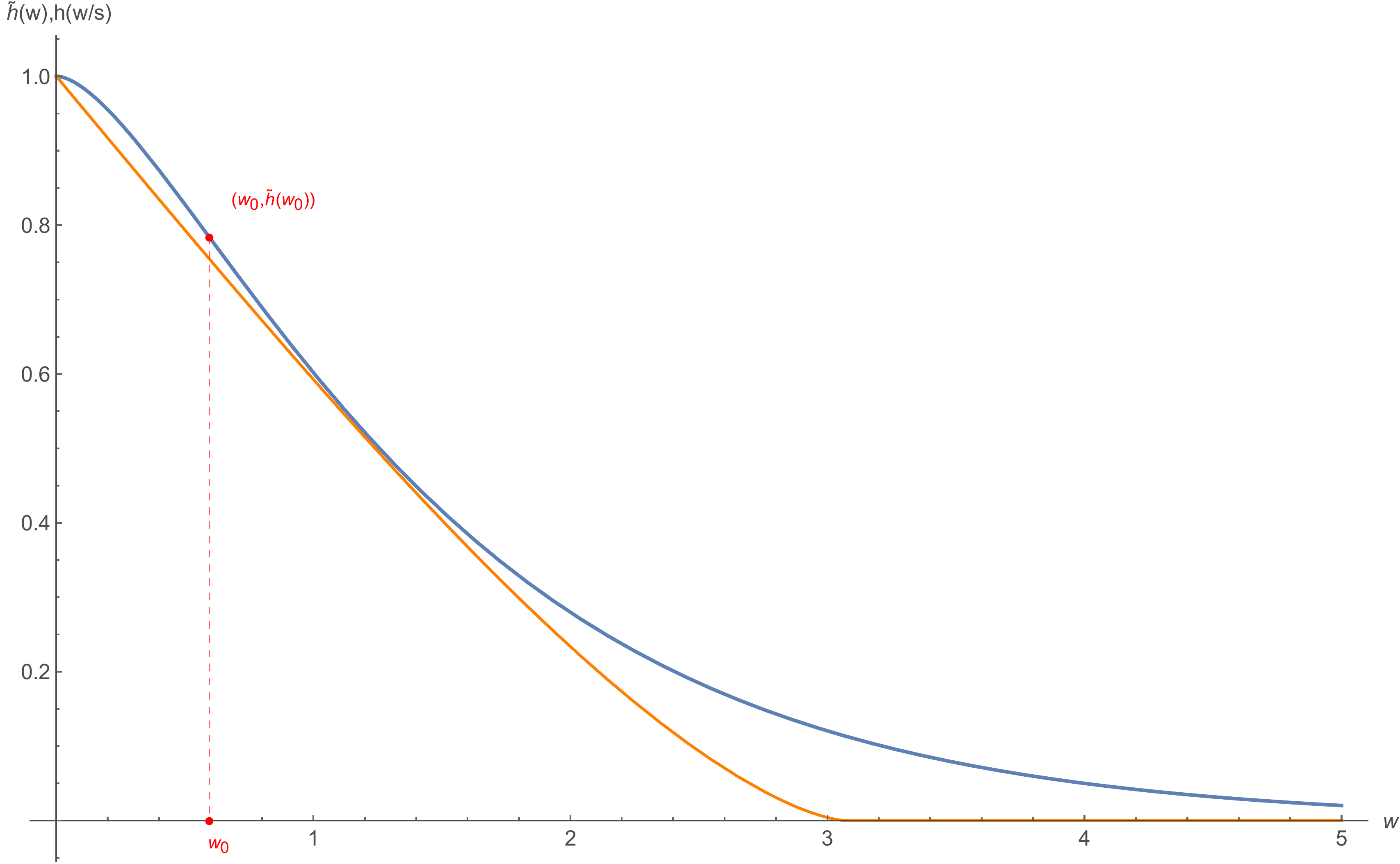} 
\centering
\caption{Plot of $h(w/s)$ scaled by $s=3.07$ and $\tilde{h}(w)$ together. }
\label{fig4}
\end{figure}

\begin{figure}[th]
\centering
\includegraphics[scale=0.45]{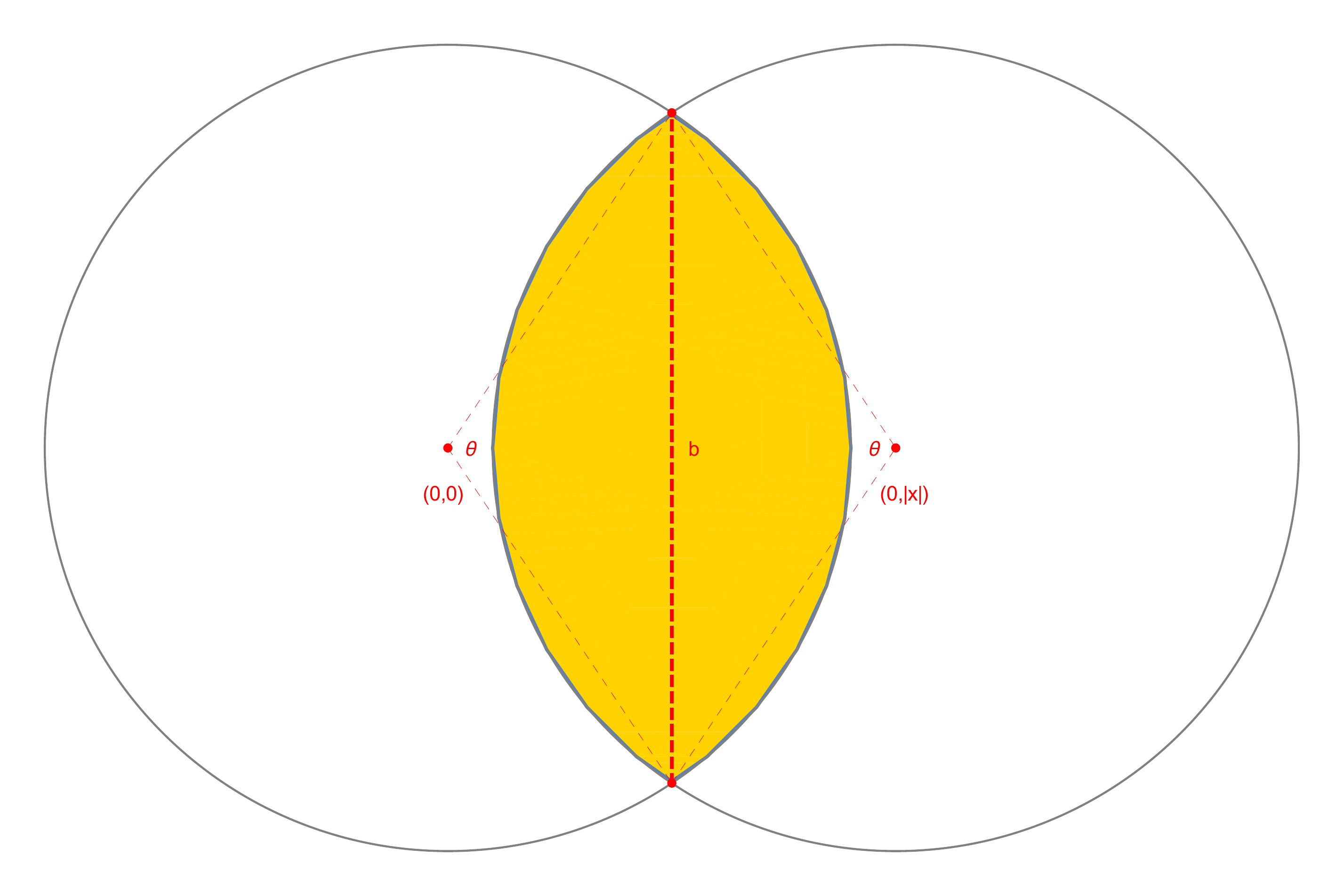} 
\centering
\caption{Geometric interpretation of the Euclid's hat function. }
\label{fig5}
\end{figure}

\noindent \textit{Proof.} We shall deduce (\ref{hw}) from (\ref{hs}) by
means of a geometric representation of the convolution integral 
\begin{equation}
\frac{\pi s^{2}}{4}h(w)=\int_{\mathbb{R}^{2}}\chi _{\lbrack 0,s/2]}(x-y)\chi
_{\lbrack 0,s/2]}(y)dy\ \qquad  \label{s2h}
\end{equation}%
(see e.g Sec. 2 of \cite{Gneiting}). The product of indicator functions does
not vanish if their support, the discs $B_{s/2}(x)$ and $B_{s/2}(0)$
centered at $x$ and $0$, intersect and this occurs when the distance $
\left\vert x\right\vert $ between their centers is less than their
diameter $s$. Writing $w=\left\vert x\right\vert /s$, we have  
\begin{equation*}
h(w)\neq 0\Longleftrightarrow 0\leq w<1~.
\end{equation*}%
From this point of view, the convolution integral (\ref{s2h}) is given by
the area $A(\theta )$ of two "caps", of common bases, made of a sector of
opening angle $\theta $ and radius $s/2$ with the triangular region inside
removed (see Fig. \ref{fig5}):%
\begin{equation}
\frac{\pi s^{2}}{4}h(w)=A(\theta )=2\times \left( \frac{1}{2}\left( \frac{s}{%
2}\right) ^{2}\theta -\frac{1}{2}\left( \frac{s}{2}\right) ^{2}\sin \theta
\right) ~,  \label{Atheta}
\end{equation}%
where, with $b$ the length of the caps common bases,%
\begin{eqnarray}
\left\vert x\right\vert &=&s\cos \theta /2  \notag \\
b &=&s\sin \theta /2~.  \label{theta}
\end{eqnarray}%
By $\left\vert x\right\vert ^{2}+b^{2}=s^{2}$, we deduce $b=s\sqrt{1-w^{2}}$%
. Solving equations (\ref{theta}) for $\theta $ and $\sin \theta $: 
\begin{eqnarray*}
\theta &=&2\arccos w \\
\sin \theta &=&2\sin \theta /2~\cos \theta /2=2w\sqrt{1-w^{2}}~,
\end{eqnarray*}%
together with (\ref{Atheta}), yields%
\begin{equation}
A(\theta )=\frac{s^{2}}{4}\left( \theta -\sin \theta \right) =\frac{s^{2}}{2}%
\left( \arccos w-w\sqrt{1-w^{2}}\right) ~.  \label{A}
\end{equation}

Equation (\ref{hw}) follows from (\ref{s2h}), (\ref{Atheta}) and (\ref{A}).
The regularity of $h(w)$ in $\left( 0,1\right) $ follows from this
representation. Since%
\begin{eqnarray*}
h^{\prime }(w) &=&-2\sqrt{1-w^{2}}<0\ , \\
h^{\prime \prime }(w) &=&\frac{2w}{\sqrt{1-w^{2}}}>0\ ,
\end{eqnarray*}%
for any $w\in \left( 0,1\right) $, we conclude that $h(w)$ is such that $%
h(0)=1$, by definition, is strictly decreasing in $\left( 0,1\right) $
monotone non increasing and convex in $\left( 0,\infty \right) $, concluding
the proof.

\hfill $\Box $

Before we state and prove our main theorem, we use (\ref{hs}) to write%
\begin{equation*}
\sum_{1\leq i,j\leq n}\sigma _{i}\sigma _{j}h\left( \left\vert
x_{i}-x_{j}\right\vert /s\right) =\frac{4}{\pi s^{2}}\int_{\mathbb{R}%
^{2}}\sum_{1\leq i,j\leq n}\sigma _{i}\sigma _{j}\chi _{\lbrack
0,s/2]}(x_{i}-x_{j}-y)\chi _{\lbrack 0,s/2]}(y)dy~.\ 
\end{equation*}%
Changing the integration variables for each term of the sum to $z=y+x_{j}$
yields%
\begin{eqnarray}
\sum_{1\leq i,j\leq n}\sigma _{i}\sigma _{j}h\left( \left\vert
x_{i}-x_{j}\right\vert /s\right) &=&\frac{4}{\pi s^{2}}\int_{\mathbb{R}%
^{2}}\sum_{1\leq i,j\leq n}\sigma _{i}\sigma _{j}\chi _{\lbrack
0,s/2]}(x_{i}-z)\chi _{\lbrack 0,s/2]}(z-x_{j})dz  \notag \\
&=&\frac{4}{\pi s^{2}}\int_{\mathbb{R}^{2}}\left( \sum_{j=1}^{n}\sigma
_{j}\chi _{\lbrack 0,s/2]}(z-x_{j})\right) ^{2}dz  \label{ssh}
\end{eqnarray}%
since the function $\chi _{\lbrack 0,s/2]}(x)$ is even.

\begin{theorem}
\label{main}For any integer $n\geq 2$, any configuration of $n$--particle $%
\left( \zeta _{1},\ldots ,\zeta _{n}\right) $, $\zeta _{j}=\left( \sigma
_{j},x_{j}\right) \in \left\{ -1,1\right\} \times \mathbb{R}^{2}$ any $s\in 
\mathbb{R}_{+}$, the total energy with interacting potential $h$ satisfies%
\begin{equation}
U_{n}(\zeta _{1},\ldots ,\zeta _{n};h(\cdot /s))=\sum_{1\leq i<j\leq
n}\sigma _{i}\sigma _{j}h\left( \left\vert x_{i}-x_{j}\right\vert /s\right)
\geq -\frac{1}{2}\left( n-\left\vert \sum\nolimits_{j=1}^{n}\sigma
_{j}\right\vert \right) ~.  \label{Unh}
\end{equation}
\end{theorem}

\noindent \textit{Proof.} Since $h(0)=1$, we add $n/2$ to the total energy
in order to include the $i=j$ terms into the sum in (\ref{Unh}):%
\begin{equation*}
\sum_{1\leq i<j\leq n}\sigma _{i}\sigma _{j}h\left( \left\vert
x_{i}-x_{j}\right\vert /s\right) =-\frac{1}{2}\sum_{j=1}^{n}\sigma
_{j}^{2}h(0)+\frac{1}{2}\sum_{1\leq i,j\leq n}\sigma _{i}\sigma _{j}h\left(
\left\vert x_{i}-x_{j}\right\vert /s\right) ~.
\end{equation*}%
So, the result is proven if we show that%
\begin{equation*}
\sum_{1\leq i,j\leq n}\sigma _{i}\sigma _{j}h\left( \left\vert
x_{i}-x_{j}\right\vert /s\right) \geq \left\vert
\sum\nolimits_{j=1}^{n}\sigma _{j}\right\vert ~.
\end{equation*}%
Using the fact that $\sum_{j=1}^{n}\sigma _{j}\chi _{\lbrack
0,s/2]}(z-x_{j}) $ is always an integer number, we have 
\begin{equation*}
\left( \sum_{j=1}^{n}\sigma _{j}\chi _{\lbrack 0,s/2]}(z-x_{j})\right)
^{2}\geq \left\vert \sum_{j=1}^{n}\sigma _{j}\chi _{\lbrack
0,s/2]}(z-x_{j})\right\vert
\end{equation*}%
and this, together with (\ref{ssh}), implies that%
\begin{eqnarray*}
\sum_{1\leq i,j\leq n}\sigma _{i}\sigma _{j}h\left( \left\vert
x_{i}-x_{j}\right\vert /s\right) &\geq &\frac{4}{\pi s^{2}}\int_{\mathbb{R}%
^{2}}\left\vert \sum\nolimits_{j=1}^{n}\sigma _{j}\chi _{\lbrack
0,s/2]}(z-x_{j})\right\vert dz \\
&\geq &\left\vert \frac{4}{\pi s^{2}}\int_{\mathbb{R}^{2}}\sum%
\nolimits_{j=1}^{n}\sigma _{j}\chi _{\lbrack 0,s/2]}(z-x_{j})dz\right\vert \\
&=&\left\vert \sum\nolimits_{j=1}^{n}\sigma _{j}\frac{4}{\pi s^{2}}\int_{%
\mathbb{R}^{2}}\chi _{\lbrack 0,s/2]}(z-x_{j})dz\right\vert \\
&=&\left\vert \sum\nolimits_{j=1}^{n}\sigma _{j}\right\vert ~,
\end{eqnarray*}%
concluding the proof.

\hfill $\Box $

\begin{remark}
Since the proof does not set any condition on the dimension of the Euclidean
space, Theorem \ref{main} holds for any $d\geq 2$. In this case, $h(w)$ has
to be replaced by the Euclid's hat $h_{d}(w)$ (see Sec. 2 of \cite{Gneiting}
for the proof of the statements in Proposition \ref{euclidhat} for the
corresponding $d$--dimensional Euclid's hat).
\end{remark}

Theorem \ref{main} implies the following

\begin{corollary}
\label{minimal}The minimal specific energy $e(h)$ and the minimal
constrained specific energy $\bar{e}(h)$, defined by (\ref{e}) and (\ref%
{ebar}), are both $-1/2$.
\end{corollary}

\noindent \textit{Proof.} This result follows from the definitions (\ref{env}%
) and (\ref{renv}) and the inequality (\ref{Unh}). The minimal specific
energy $e(h)=\inf_{n\geq 2}e_{n}(h)$ of $h$ is attained for even number of
particles $n$ satisfying $\displaystyle\sum\nolimits_{j=1}^{n}\sigma _{j}=0$
and $\displaystyle\sum\nolimits_{j=1}^{n}\sigma _{j}^{2}=n$ when they
collapse to a single point since, in this case, the inequality (\ref{Unh})
becomes an equality. Likewise, the constrained minimal specific energy $\bar{%
e}(h)=\inf_{n\geq 2}\bar{e}_{n}(h)$ of $h$ is attained for odd number of
particles $n$ satisfying $\left\vert \displaystyle\sum\nolimits_{j=1}^{n}%
\sigma _{j}\right\vert =1$ and $\displaystyle\sum\nolimits_{j=1}^{n}\sigma
_{j}^{2}=n$ when they collapse to a single point. Note that, for a
calculation similar to the energy in (\ref{delta}), the potential energy (%
\ref{Unzetas}) with $n=2r+1$, $\sigma _{1}=\cdots =\sigma _{r}=-\sigma
_{r+1}=\cdots =-\sigma _{2r+1}$ and $x_{1}=\cdots =x_{2r+1}=x_{0}\in \mathbb{%
R}^{2}$, is given by ($h(0)=1$)%
\begin{equation*}
U_{n}(\zeta _{1},\ldots ,\zeta _{n};h)=-(r+1)r+\frac{r(r-1)}{2}+\frac{(r+1)r%
}{2}=-r=\frac{-1}{2}(n-1)~.
\end{equation*}

\hfill $\Box $

As a consequence of Corollary \ref{minimal}, representing the Yukawa
potential $v$ as scale mixtures of Euclid's
hat (\ref{vt0t}) regularized at short distances, the stability bound
(\ref{stability}) can be replaced by (\ref{stability1}) with 
\begin{equation*}
B=\bar{B}=\frac{1}{2}\int_{t_{0}}^{t}g(s)ds\ .
\end{equation*}

\paragraph{Properties of the mixture function}

Regarding the mixture function, we have the following

\begin{proposition}
\label{exponential}The function $g:\left( 0,\infty \right) \longrightarrow
\left( 0,\infty \right) $ given by (\ref{gs}) can be written as 
\begin{equation*}
g(s)=\frac{1}{2\pi s}m(s)
\end{equation*}%
where%
\begin{equation}
m(s)=\frac{1}{2}\int_{s}^{\infty }y^{2}K_{1}(y)\frac{y}{\sqrt{y^{2}-s^{2}}}dy
\label{m}
\end{equation}%
is a regular function such that $m(0)=1$, increases monotonously in $\left(
0,s_{0}\right) $, where $s_{0}=0.812$(\ldots ) and $m(s_{0})=m_{0}=1,075$%
(\ldots ), then decreses monotonously in $\left( s_{0},\infty \right) $ to $%
0 $, exponentially fast. Globally, it is bounded from above and from below
as 
\begin{equation}
\frac{\pi }{4}e^{-s}(1+s+s^{2})<m(s)<\frac{\pi }{4}e^{-s}(3+3s+s^{2})~,%
\qquad \forall ~s\in \lbrack 0,\infty )~.  \label{lubound}
\end{equation}%
In the vicinity of the origin, it satisfies 
\begin{equation}
m(s)\leq 1+\left( a-\frac{1}{4}\log s\right) s^{2},\qquad s\in \left[ 0,1%
\right]  \label{ms}
\end{equation}%
where $a=(1-3\gamma +\log 4-\psi (-1/2))/8=0.07726$(\ldots ), being the
r.h.s. of (\ref{ms}) asymptotic to $m(s)$ at $s=0$.
\end{proposition}

\begin{figure}[th]
\centering
\includegraphics[scale=0.45]{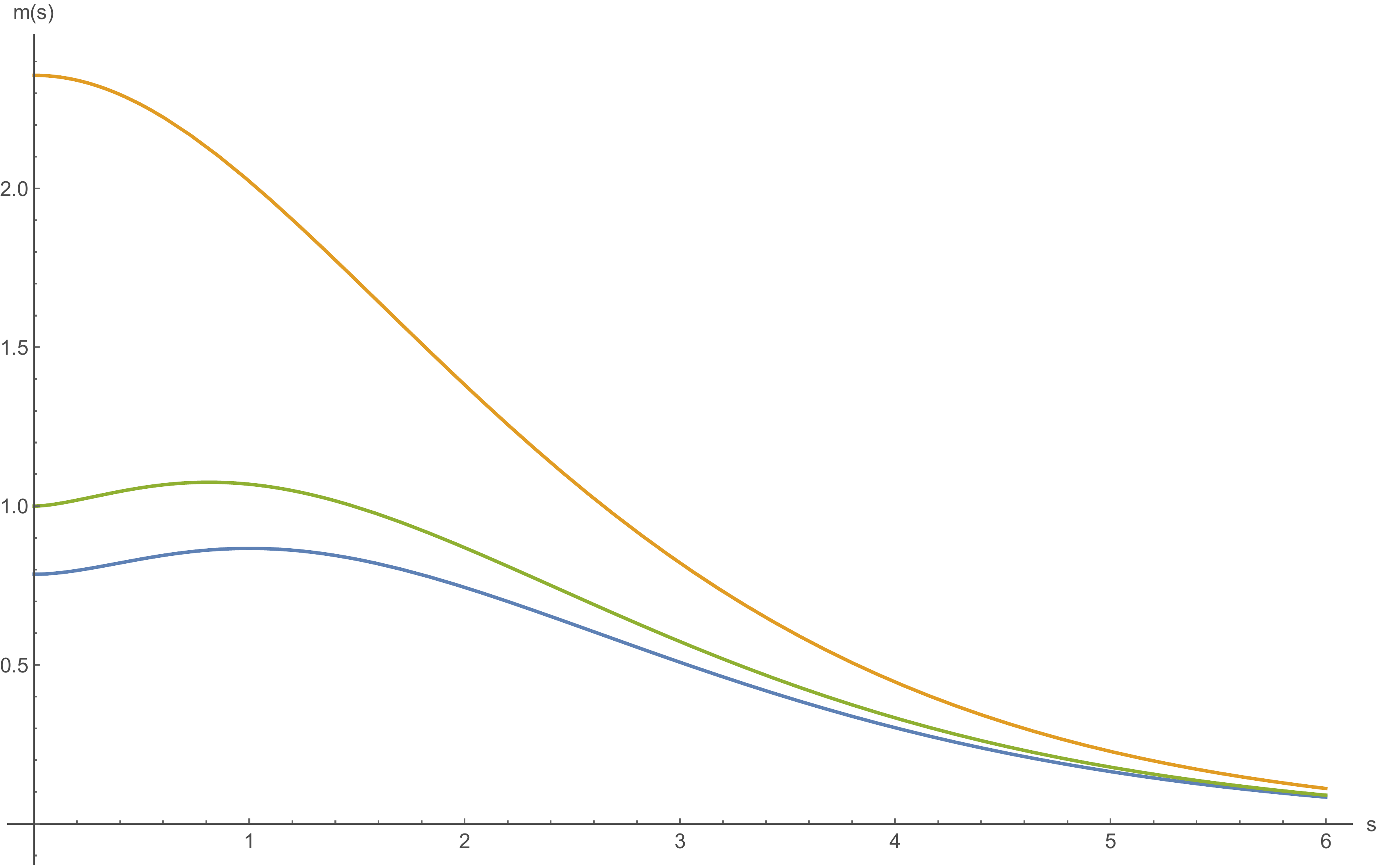} 
\centering
\caption{Plot of $m(s)$ together with its upper and lower functions. }
\label{fig6}
\end{figure}

\noindent \textit{Proof.} We begin by showing that (\ref{gs}) multiplied by $%
2\pi s$ can be written as (\ref{m}). For this, we use $K_{1}(w)=-K_{0}^{%
\prime }(w)$ and the representation (see \cite{Glimm-Jaffe}, Section 7.2) 
\begin{equation}
K_{0}(w)=\int_{0}^{\infty }e^{-w\sqrt{k^{2}+1}}\frac{dk}{\sqrt{k^{2}+1}}~,
\label{K0}
\end{equation}%
from which we infer that $K_{0}$ is regular in $\left( 0,\infty \right) $.
We may thus differentiate (\ref{K0}) three times, replace it into (\ref{gs}%
), exchange the integration order and, after multiplying by $2\pi s$ it can
be written as 
\begin{equation}
m(s)=\int_{0}^{\infty }\left( k^{2}+1\right) F(s,k)dk  \label{mF}
\end{equation}%
where%
\begin{eqnarray}
F(s,k) &=&\frac{s^{2}}{2}\int_{s}^{\infty }e^{-r\sqrt{k^{2}+1}}\frac{rdr}{%
\sqrt{r^{2}-s^{2}}}  \notag \\
&=&\frac{s^{3}}{2}\int_{0}^{\infty }e^{-s\sqrt{k^{2}+1}\sqrt{z^{2}+1}}dz 
\notag \\
&=&-\frac{s^{3}}{2}K_{0}^{\prime }(s\sqrt{k^{2}+1})  \notag \\
&=&\frac{s^{3}}{2}K_{1}(s\sqrt{k^{2}+1})~.  \label{F}
\end{eqnarray}%
We have changed variable $sz=\sqrt{r^{2}-s^{2}}$, so $r=s\sqrt{z^{2}+1}$ and 
$rdr/\sqrt{r^{2}-s^{2}}=sdz$. Replacing (\ref{F}) back into (\ref{mF}),
making one more change of variable: $s\sqrt{k^{2}+1}=y$, so that $sk=\sqrt{%
y^{2}-s^{2}}$ and $sdk=y/\sqrt{y^{2}-s^{2}}$, yields (\ref{m}).

The sequence of operations bringing (\ref{gs}) into the form (\ref{m}) will
be applied some more times. Let us start by finding a lower bound for (\ref%
{m}). By monotonicity of the modified Bessel functions with respect to their
order (see \cite{Cochran}) and integration by parts, we have 
\begin{eqnarray}
m(s) &>&\frac{1}{2}\int_{s}^{\infty }y^{2}K_{0}(y)\frac{y}{\sqrt{y^{2}-s^{2}}%
}dy  \notag \\
&=&\frac{-1}{2}\int_{s}^{\infty }\left( y^{2}K_{0}(y)\right) ^{\prime }\sqrt{%
y^{2}-s^{2}}dy  \notag \\
&=&L(s)-J(s)  \label{K-J}
\end{eqnarray}%
where%
\begin{eqnarray}
L(s) &=&\frac{1}{2}\int_{s}^{\infty }y^{2}K_{1}(y)\sqrt{y^{2}-s^{2}}dy
\label{K} \\
J(s) &=&\int_{s}^{\infty }yK_{0}(y)\sqrt{y^{2}-s^{2}}dy~.  \label{J}
\end{eqnarray}%
Observe that the boundary term in the partial integration, $\left.
y^{2}K_{0}(y)\sqrt{y^{2}-s^{2}}/2\right\vert _{y=s}^{\infty }$ vanishes for
all $s\in \left( 0,\infty \right) $ because the exponential decay of $%
K_{0}(y)$ and boundedness of $y^{2}K_{0}(y)$.

\begin{lemma}
\label{IIs}Let $I:[0,\infty )\longrightarrow \mathbb{R}$ be defined by 
\begin{equation}
I(s)=\int_{s}^{\infty }K_{1}(y)\sqrt{y^{2}-s^{2}}dy~.  \label{I}
\end{equation}%
The integral can be written as%
\begin{equation}
I(s)=\int_{s}^{\infty }K_{1}(y)\frac{s}{\sqrt{y^{2}-s^{2}}}dy  \label{Is}
\end{equation}%
and from these we conclude that $I(s)=\pi e^{-s}/2$.
\end{lemma}

\noindent \textit{Proof.} Since the integral (\ref{I}) converge uniformly in 
$[s_{0},K)$, for any $s_{0}>0$ and $K<\infty $, the integral (\ref{Is}) is
minus the derivative of the integral (\ref{I}):%
\begin{equation}
I^{\prime }(s)=\int_{s}^{\infty }K_{1}(y)\frac{-s}{\sqrt{y^{2}-s^{2}}}%
dy=-I(s)~,\qquad s>0~.  \label{II}
\end{equation}%
Observe that $\left. K_{1}(y)\sqrt{y^{2}-s^{2}}\right\vert _{y=s}=0$ for the
same reason as before. Since $ae^{-s}$ solves (\ref{II}) for any $a\in 
\mathbb{R}$, the proof will be completed once we establish that (\ref{I})
implies (\ref{Is}) and show $I(0)=\pi /2$. We begin with the latter.

Repeating the operations bringing (\ref{gs}) into the form (\ref{m}), it
follows from (\ref{K0}) and $K_{0}^{\prime }(y)=-K_{1}(y)$ that 
\begin{eqnarray}
I(0) &=&\int_{0}^{\infty }yK_{1}(y)dy  \notag \\
&=&\int_{0}^{\infty }y\left( \int_{0}^{\infty }e^{-y\sqrt{k^{2}+1}}dk\right)
dy  \notag \\
&=&\int_{0}^{\infty }\left( \int_{0}^{\infty }ye^{-y\sqrt{k^{2}+1}}dy\right)
dk  \notag \\
&=&\int_{0}^{\infty }\frac{-1}{\sqrt{k^{2}+1}}\left( \left. ye^{-y\sqrt{%
k^{2}+1}}\right\vert _{y=0}^{\infty }-\int_{0}^{\infty }e^{-y\sqrt{k^{2}+1}%
}dy\right) dk  \notag \\
&=&\int_{0}^{\infty }\frac{1}{k^{2}+1}dk=\left. \arctan k\right\vert
_{k=0}^{\infty }=\frac{\pi }{2}~.  \label{I0}
\end{eqnarray}

Now, we develop (\ref{I}) as

\begin{eqnarray*}
\int_{s}^{\infty }K_{1}(y)\sqrt{y^{2}-s^{2}}dy &=&\int_{s}^{\infty }\left(
\int_{0}^{\infty }e^{-y\sqrt{k^{2}+1}}dk\right) \sqrt{y^{2}-s^{2}}dy \\
&=&\int_{0}^{\infty }\left( \int_{s}^{\infty }e^{-y\sqrt{k^{2}+1}}\sqrt{%
y^{2}-s^{2}}dy\right) dk \\
&=&\int_{0}^{\infty }\left( \int_{s}^{\infty }e^{-y\sqrt{k^{2}+1}}\frac{y}{%
\sqrt{y^{2}-s^{2}}}dy\right) \frac{dk}{\sqrt{k^{2}+1}} \\
&=&s\int_{0}^{\infty }\left( \int_{0}^{\infty }e^{-s\sqrt{k^{2}+1}\sqrt{%
r^{2}+1}}dr\right) \frac{dk}{\sqrt{k^{2}+1}} \\
&=&s\int_{0}^{\infty }K_{1}\left( s\sqrt{k^{2}+1}\right) \frac{dk}{\sqrt{%
k^{2}+1}} \\
&=&s\int_{s}^{\infty }K_{1}\left( y\right) \frac{dy}{\sqrt{y^{2}-s^{2}}}
\end{eqnarray*}%
In the second equality we exchange the integration order, then we integrate
by parts; we change variable $y=s\sqrt{r^{2}+1}$ in the fourth equality, use
(\ref{K0}) together with $K_{0}^{\prime }(w)=-K_{1}(w)$ and in the last
equality we change again the variable $y=s\sqrt{k^{2}+1}$. This concludes
the proof of the lemma.

\hfill $\Box $

Returning to the proof of Proposition \ref{exponential}, we now deduce an
equation for $J(s)$ and $L(s)$ in terms of $I(s)$. Differentiating (\ref{J})
with respect to $s$, gives%
\begin{eqnarray}
J^{\prime }(s) &=&-\left. yK_{0}(y)\sqrt{y^{2}-s^{2}}\right\vert
_{y=s}-\int_{s}^{\infty }yK_{0}(y)\frac{s}{\sqrt{y^{2}-s^{2}}}dy  \notag \\
&=&-s\int_{s}^{\infty }K_{0}(y)\left( \sqrt{y^{2}-s^{2}}\right) ^{\prime }dy
\notag \\
&=&-\left. sK_{0}(y)\sqrt{y^{2}-s^{2}}\right\vert _{y=s}^{\infty
}-s\int_{s}^{\infty }K_{1}(y)\sqrt{y^{2}-s^{2}}dy  \notag \\
&=&-sI(s)=\frac{-\pi }{2}se^{-s}~,  \label{Jlinha}
\end{eqnarray}%
by Lemma \ref{IIs}. Analogously, differentiating (\ref{K}) with respect to $%
s $, together with (\ref{xnKn}), gives%
\begin{eqnarray}
L^{\prime }(s) &=&\frac{-s}{2}\int_{s}^{\infty }yK_{1}(y)\frac{y}{\sqrt{%
y^{2}-s^{2}}}dy  \notag \\
&=&\frac{s}{2}\int_{s}^{\infty }\left( yK_{1}\right) ^{\prime }(y)\sqrt{%
y^{2}-s^{2}}dy  \notag \\
&=&\frac{-s}{2}\int_{s}^{\infty }yK_{0}(y)\sqrt{y^{2}-s^{2}}dy  \notag \\
&=&-\frac{1}{2}sJ(s)\ .  \label{Klinha}
\end{eqnarray}%
We need also initial condition to both equations. Performing as in (\ref{I0}%
),%
\begin{eqnarray}
J(0) &=&\int_{0}^{\infty }y^{2}K_{0}(y)dy  \notag \\
&=&\int_{0}^{\infty }y^{2}\left( \int_{0}^{\infty }e^{-y\sqrt{k^{2}+1}}\frac{%
dk}{\sqrt{k^{2}+1}}\right) dy  \notag \\
&=&\int_{0}^{\infty }\left( \int_{0}^{\infty }y^{2}e^{-y\sqrt{k^{2}+1}%
}dy\right) \frac{dk}{\sqrt{k^{2}+1}}  \notag \\
&=&2\int_{0}^{\infty }\frac{dk}{\left( k^{2}+1\right) ^{2}}  \notag \\
&=&\left. \left( \arctan k+\frac{k}{k^{2}+1}\right) \right\vert
_{k=0}^{\infty }=\frac{\pi }{2}~  \label{J0}
\end{eqnarray}%
and%
\begin{eqnarray}
L(0) &=&\frac{1}{2}\int_{0}^{\infty }y^{3}K_{1}(y)dy  \notag \\
&=&\frac{1}{2}\int_{0}^{\infty }y^{3}\left( \int_{0}^{\infty }e^{-y\sqrt{%
k^{2}+1}}dk\right) dy  \notag \\
&=&\frac{1}{2}\int_{0}^{\infty }\left( \int_{0}^{\infty }y^{3}e^{-y\sqrt{%
k^{2}+1}}dy\right) dk  \notag \\
&=&3\int_{0}^{\infty }\frac{dk}{\left( k^{2}+1\right) ^{2}}=\frac{3\pi }{4}~.
\label{K00}
\end{eqnarray}%
Integrating (\ref{Jlinha}) together with (\ref{J0}), yields%
\begin{eqnarray}
J(s) &=&J(0)-\frac{\pi }{2}\int_{0}^{s}te^{-t}dt  \notag \\
&=&\frac{\pi }{2}\left( 1+se^{-s}-\int_{0}^{s}e^{-t}dt\right) =\frac{\pi }{2}%
(1+s)e^{-s}~.  \label{Js}
\end{eqnarray}%
Analogously, integrating (\ref{Klinha}) together with (\ref{Js}) and (\ref%
{K00}), yields%
\begin{eqnarray}
L(s) &=&L(0)-\frac{\pi }{4}\int_{0}^{s}\left( 1+t\right) te^{-t}dt  \notag \\
&=&\frac{\pi }{4}\left( 3+\left( 1+s\right) se^{-s}-\int_{0}^{s}\left(
1+2t\right) e^{-t}dt\right)  \notag \\
&=&\frac{\pi }{4}\left( 2+\left( 1+s\right) se^{-s}+\left( 1+2s\right)
e^{-s}-2\int_{0}^{s}e^{-t}dt\right)  \notag \\
&=&\frac{\pi }{4}\left( 3+3s+s^{2}\right) e^{-s}~.  \label{Ks}
\end{eqnarray}%
Equations (\ref{Js}) and (\ref{Ks}) replaced into (\ref{K-J}) gives the
lower bound (\ref{lubound}).

An upper bound is obtained similarly. By monotonicity of the modified Bessel
functions with respect to their order (see \cite{Cochran}) and integration
by parts, we have 
\begin{eqnarray}
m(s) &<&\frac{1}{2}\int_{s}^{\infty }y^{2}K_{2}(y)\frac{y}{\sqrt{y^{2}-s^{2}}%
}dy  \notag \\
&=&\frac{-1}{2}\int_{s}^{\infty }\left( y^{2}K_{2}\right) ^{\prime }(y)\sqrt{%
y^{2}-s^{2}}dy=L(s)  \label{mK}
\end{eqnarray}%
by (\ref{xnKn}), where $L(s)$ is given by (\ref{K}). Equation (\ref{mK})
together with (\ref{Ks}) gives the upper bound (\ref{lubound}).

\begin{figure}[th]
\centering
\includegraphics[scale=0.45]{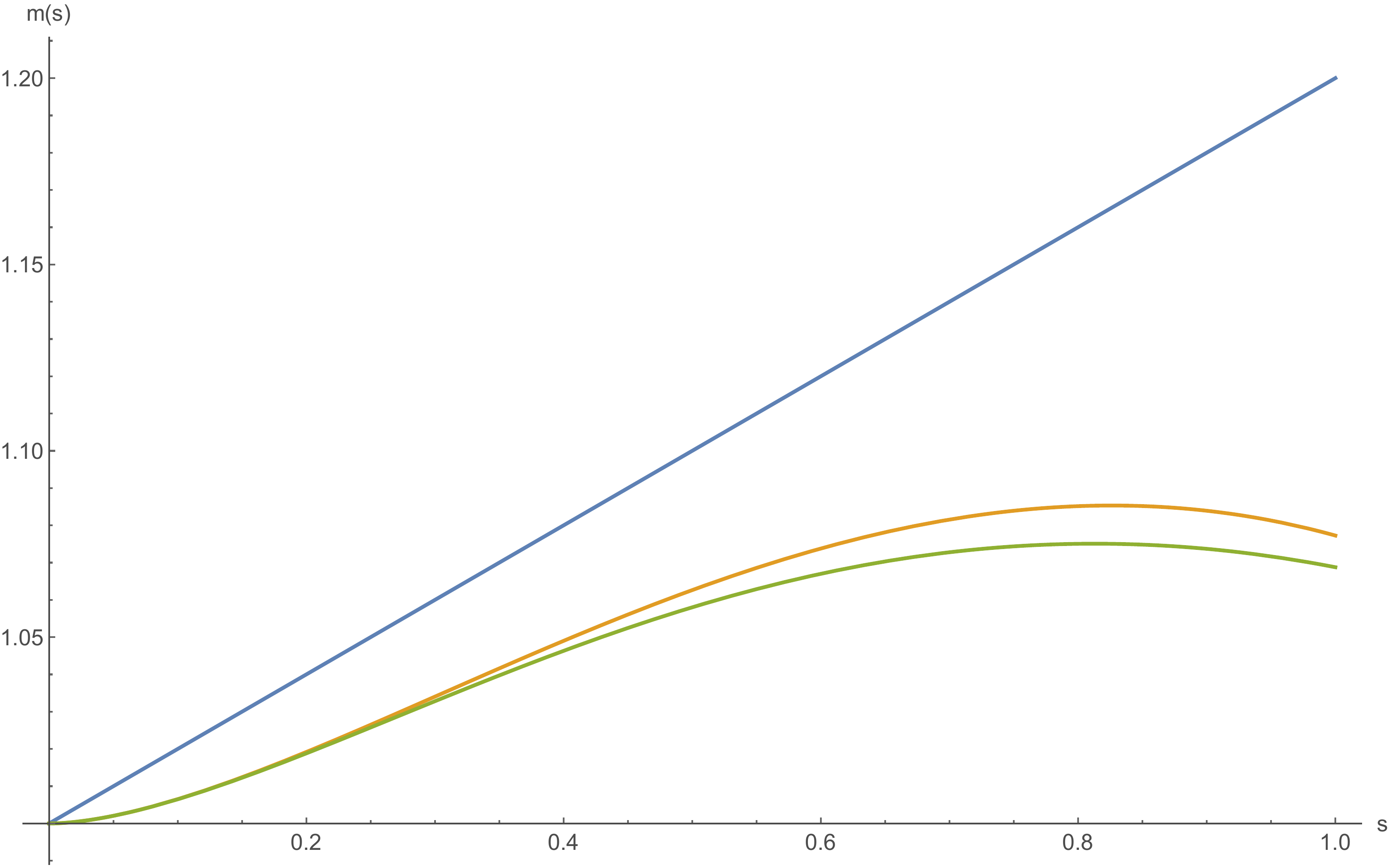} 
\centering
\caption{Plot of $m(s)$ together with its best
and linear (upper) asymptotes.  }
\label{fig7}
\end{figure}

The asymptotic behavior (\ref{ms}) of $m(s)$ follows from the mean value
theorem%
\begin{equation}
m(s)-m(0)=\int_{0}^{s}m^{\prime }(t)dt=m^{\prime }(\tilde{s})s  \label{mm}
\end{equation}%
for some $\tilde{s}=\tilde{s}(s)\in \lbrack 0,s]$ depending on $s$. The
value $m(0)$ may be calculated as $I(0)$ in (\ref{I0}), using the
representation (\ref{K0}) for $K_{1}(y)=-K_{0}^{\prime }(y)$ and exchange
the integration order:%
\begin{eqnarray*}
m(0) &=&\int_{0}^{\infty }y^{2}K_{1}(y)dy \\
&=&\int_{0}^{\infty }\left( \int_{0}^{\infty }y^{2}e^{-y\sqrt{k^{2}+1}%
}dy\right) dk \\
&=&-\int_{0}^{\infty }\frac{1}{\left( k^{2}+1\right) ^{3/2}}dk=\left. \frac{k%
}{\sqrt{k^{2}+1}}\right\vert _{0}^{\infty }=1~.
\end{eqnarray*}%
To calculate the derivative of $m(s)$ we apply integration by parts twice,
before and after the derivative with respect to $s$:%
\begin{eqnarray*}
m(s) &=&\frac{-1}{2}\int_{s}^{\infty }\left( y^{2}K_{1}(y)\right) ^{\prime }%
\sqrt{y^{2}-s^{2}}dy \\
&=&\frac{-1}{2}\int_{s}^{\infty }\left( yK_{1}(y)-y^{2}K_{0}(y)\right) \sqrt{%
y^{2}-s^{2}}dy
\end{eqnarray*}%
by $\left( y\cdot \left( yK_{1}\right) \right) ^{\prime }=yK_{1}+y\left(
yK_{1}\right) ^{\prime }$ together with (\ref{xnKn}); by $%
K_{1}+yK_{1}^{\prime }=\left( yK_{1}\right) ^{\prime }=-yK_{0}$ we have $%
-K_{1}^{\prime }=K_{0}+K_{1}/y$ and 
\begin{eqnarray}
m^{\prime }(s) &=&\frac{s}{2}\int_{s}^{\infty }\left(
K_{1}(y)-yK_{0}(y)\right) \frac{y}{\sqrt{y^{2}-s^{2}}}dy  \notag \\
&=&\frac{-s}{2}\int_{s}^{\infty }\left( K_{1}(y)-yK_{0}(y)\right) ^{\prime }%
\sqrt{y^{2}-s^{2}}dy  \notag \\
&=&M(s)+\frac{s}{2}N(s)  \label{MN}
\end{eqnarray}%
where%
\begin{eqnarray}
M(s) &=&\frac{s}{2}\int_{s}^{\infty }K_{1}(y)\frac{\sqrt{y^{2}-s^{2}}}{y}dy 
\notag \\
&\leq &\frac{s}{2}\int_{s}^{\infty }K_{1}(y)dy=\frac{s}{2}K_{0}(s)  \label{M}
\end{eqnarray}%
in view of the inequality $\sqrt{y^{2}-s^{2}}/y\leq 1$ for $s\leq y<\infty $%
, $-K_{0}^{\prime }(y)=K_{1}(y)>0$ and the fundamental theorem of calculus.
Both boundary terms yielded from the partial integrations vanish. We observe
that%
\begin{equation*}
K_{0}(s)=-\left. \frac{\partial I_{\nu }(s)}{\partial \nu }\right\vert _{\nu
=0}=-\log \left( s/2\right) \sum_{n=0}^{\infty }\frac{\left( s/2\right) ^{2n}%
}{\left( n!\right) ^{2}}+\sum_{n=0}^{\infty }\frac{\left( s/2\right) ^{2n}}{%
\left( n!\right) ^{2}}\psi (1+n)
\end{equation*}%
where $\psi (z)=\Gamma ^{\prime }(z)/\Gamma (z)$ is the digamma function and
so, $K_{0}(s)=-\log (s/2)-\gamma +O(s^{2})$ where $\gamma =-\psi (1)$ is the
Euler-Mascheroni constant. The other term of (\ref{MN}) can analogously be
bounded by 
\begin{eqnarray}
N(s) &=&\int_{s}^{\infty }\left( 2yK_{0}(y)-y^{2}K_{1}(y)\right) \frac{\sqrt{%
y^{2}-s^{2}}}{y}dy  \notag \\
&\leq &\int_{s}^{\infty }\left( 2yK_{0}(y)-y^{2}K_{1}(y)\right) dy\leq 0~,
\label{N}
\end{eqnarray}%
provided $s\in \left[ 0,\bar{s}\right] $ where $\bar{s}\geq 3/2$. For this,
we used that $\sqrt{y^{2}-s^{2}}/y\leq 1$ and by (\ref{KK}), 
\begin{equation}
2yK_{0}(y)-y^{2}K_{1}(y)=yK_{0}(y)\left( 2-\frac{yK_{1}(y)}{K_{0}(y)}\right)
\geq yK_{0}(y)\left( 2-x-\frac{1}{2}\right) \geq 0  \label{integrand}
\end{equation}%
if $x\leq 3/2$. Consequently, for $s\leq 3/2$ the integral of the l.h.s. of (%
\ref{integrand}) over $[0,s)$ is positive and, using the representation (\ref%
{K0}) for $K_{0}(y)$ and for $K_{1}(y)=-K_{0}^{\prime }(y)$ and exchange the
integration order, we have 
\begin{eqnarray*}
\int_{s}^{\infty }\left( 2yK_{0}(y)-y^{2}K_{1}(y)\right) dy &\leq
&\int_{0}^{\infty }\left( 2yK_{0}(y)-y^{2}K_{1}(y)\right) dy \\
&=&\int_{0}^{\infty }\left( \int_{0}^{\infty }\left( 2y-y^{2}\sqrt{k^{2}+1}%
\right) e^{-y\sqrt{k^{2}+1}}dy\right) \frac{dk}{\sqrt{k^{2}+1}} \\
&=&\int_{0}^{\infty }\left( \int_{0}^{\infty }\left( 2y-2y\right) e^{-y\sqrt{%
k^{2}+1}}dy\right) \frac{dk}{\sqrt{k^{2}+1}}=0
\end{eqnarray*}%
by integration by parts. To obtain (\ref{ms}) and conclude the proof of
Proposition \ref{exponential}, we need to optimize the choice of $\tilde{s}%
(s)$ in (\ref{mm}). So far, by (\ref{mm}), (\ref{MN}), (\ref{M}) and (\ref{N}%
) we have%
\begin{eqnarray}
m(s) &\leq &1+\frac{1}{2}\int_{0}^{s}tK_{0}(t)ds  \notag \\
&=&1+\frac{1}{2}(1-sK_{1}(s))  \label{mh}
\end{eqnarray}%
by Proposition \ref{h1} and this upper bound is asymptotic as $s$ tends to $%
0 $: $m(s)=1+O\left( s^{2}\right) $ the $s^{2}$ order term in the upper
bound is $(1-2\gamma -2\log (s/2))/8=0.1539$(\ldots )$~-(\log s)/4$. The
best upper bound up to $O\left( s^{2}\right) $ term is, however, stated in
Proposition \ref{exponential}, given by the asymptotic expansion of
(\ref{g-s}), calculated algebraically by the software Mathematica. 

\hfill $\Box $

\section{Majorant of the density function\label{MDF}}

\setcounter{equation}{0} \setcounter{theorem}{0}

\paragraph{Set up and ingredients}

Let $\left( \Omega ,\mathcal{B},\varrho \right) $ denote the (translational
invariant) $\sigma $--finite measure space on $\left\{ -1,1\right\} \times 
\mathbb{R}^{2}$; the set $\Omega $ corresponds to the possible
configurations of a single particle (we united $\sigma $ and $x$ into $\zeta
=\left( \sigma ,x\right) $) and $\displaystyle\int d\varrho (\zeta )\cdot
~=1/2\displaystyle\sum_{\sigma \in \{-1,1\}}\displaystyle\int_{\mathbb{R}%
^{2}}d^{2}x~\cdot ~$denotes the integration with respect to $\rho $. Let%
\begin{equation}
\beta p(\beta ,z)=\sum_{n\geq 1}\frac{z^{n}}{n!}\int d\varrho (\zeta
_{2})\cdots d\varrho (\zeta _{n})\psi _{n}^{c}(\zeta _{1},\ldots ,\zeta
_{n};\beta v)  \label{betap}
\end{equation}%
be the pressure of the Yukawa gas in the infinite volume limit, where $v$ is
the Yukawa potential regularized at short distances $s\leq t_{0}$, given by
the scale decomposition (\ref{vtx}). We observe that, as $v$ decays
exponentially fast at infinity and has its singularity at origin removed,
the finite volume pressure $p_{\Lambda _{i}}$, defined for any increasing
sequence $\left( \Lambda _{i}\right) _{i\geq 1}$ of squares with $%
\lim_{i}\Lambda _{i}=\mathbb{R}^{2}$,\footnote{%
Given by (\ref{betap}) with the integral over the $n$--particle
configurations restricted to $\Lambda _{i}$ divided by $\left\vert \Lambda
_{i}\right\vert $: $\beta p_{\Lambda _{i}}=\displaystyle\sum_{n\geq 1}\dfrac{%
z^{n}}{n!}\dfrac{1}{\left\vert \Lambda _{i}\right\vert }\displaystyle%
\int_{\Lambda _{i}^{n}}d\varrho (\zeta _{1})\cdots d\varrho (\zeta _{n})\psi
_{n}^{c}$} converges by standard methods (see e.g. \cite{Ruelle}) and
translational invariance of $v$ to the expression (\ref{betap}).

The density function $\rho (\beta ,z)=z\partial p/\partial z(\beta ,z)$ is
another thermodynamical function which will be convenient to write as an
Mayer series (\ref{p}) in power of the activity $z$: 
\begin{equation}
\frac{\beta }{z}\rho (\beta ,z)=\sum_{n\geq 1}nb_{n}z^{n-1}  \label{rho}
\end{equation}%
where $b_{1}=1$ and, for $n>1$,%
\begin{equation*}
b_{n}=\frac{1}{n!}\int d\varrho (\zeta _{2})\cdots d\varrho (\zeta _{n})\psi
_{n}^{c}(\zeta _{1},\ldots ,\zeta _{n};\beta v)
\end{equation*}%
is the so called $n$--th Mayer coefficient in the infinite volume limit.
Note that $\beta \rho (\beta ,z)/z=1$ is the equation of state of an ideal
gas and due the interaction of the charged particles through the Yukawa pair
potential, the series (\ref{rho}) provides corrections to all order about it
expressed in terms of the Ursell (cluster) functions $\psi _{n}^{c}$. A
formal power series in $z$ 
\begin{equation*}
\Theta ^{\ast }(z)=\sum_{n\geq 1}C_{n}^{\ast }z^{n-1}
\end{equation*}%
is a majorant of $\beta \rho (\beta ,z)/z$ if the $C_{n}^{\ast }$ are
nonnegative and 
\begin{equation*}
n\left\vert b_{n}\right\vert \leq C_{n}^{\ast }
\end{equation*}%
holds for all $n\in \mathbb{N}$. It follows that, if the $\Theta ^{\ast }(z)$
series converges on the open disc $D(r):=\left\{ z\in \mathbb{C}:\left\vert
z\right\vert <r\right\} $ for some $r>0$, then $\rho (\beta ,z)$ is
holomorphic function of $z$ on the same disc. The largest $r$ provides an
lower bound on the radius of convergence of the Mayer series (\ref{rho}) and
(\ref{p}).

For the problem at our hand, the most efficient method of constructing
majorants combines (multi)scale decomposition of $v$ together with some
basic ingredients. Beginning with the latter, the following elementary
lemmas are useful.

\begin{lemma}
\label{elementary}If $a$, $b$, $c$ and $d$ are positive numbers such that $%
a-c$ and $b-d$ are positive, then $ab-cd$ is also positive.
\end{lemma}

\noindent \textit{Proof.} Writing%
\begin{eqnarray}
ab-cd &=&ab-\frac{1}{2}\left( ad+bc\right) -\left( cd-\frac{1}{2}\left(
ad+bc\right) \right)  \notag \\
&=&\frac{1}{2}\left( a (b-d)+ (a-c) b -(c-a) d - c (d-b) \right)  \notag \\
&=&\frac{1}{2}\left( \left( a+c\right) \left( b-d\right) +\left( a-c\right)
\left( b+d\right) \right) >0~,  \label{abcd}
\end{eqnarray}%
concluding the proof.

\hfill $\Box $

\begin{lemma}
\label{elementary1}Let $a=\left( a_{n}\right) _{n\geq 1}$, $b=\left(
b_{n}\right) _{n\geq 1}$, $\tilde{a}=\left( \tilde{a}_{n}\right) _{n\geq 1}$
and $\tilde{b}=\left( \tilde{b}_{n}\right) _{n\geq 1}$ be positive numerical
sequences ($a$, $b$, $\tilde{a}$ and $\tilde{b}>0$) such that $\tilde{a}-a$
and $\tilde{b}-b$ are both positive sequences (i.e., $\tilde{a}_{n}-a_{n}>0$
and $\tilde{b}_{n}-b_{n}>0$ hold for all $n\geq 1$). Let the convolution
product $e=c\ast d$ and the pointwise product $f=c\cdot d$ of two sequences $%
c=\left( c_{n}\right) _{n\geq 1}$ and $d=\left( d_{n}\right) _{n\geq 1}$ be
defined by the sequences $e=\left( e_{n}\right) _{n\geq 1}$ and $f=\left(
f_{n}\right) _{n\geq 1}$ where $e_{1}=0$ and%
\begin{equation*}
e_{n}=\sum_{k=1}^{n-1}c_{k}d_{n-k}~,\qquad n\geq 2~
\end{equation*}%
and%
\begin{equation*}
f_{n}=c_{n}d_{n}~,\qquad n\geq 1\ .
\end{equation*}%
Then, \textbf{(i)} $\tilde{a}\cdot \tilde{b}-a\cdot b>0$; \textbf{(ii)} $%
\tilde{a}\ast \tilde{b}-a\ast b>0$; in particular \textbf{(iii)} $\tilde{a}%
\ast \tilde{a}-a\ast a>0$ and $\tilde{b}\ast \tilde{b}-b\ast b>0$ hold.
\end{lemma}

\noindent \textit{Proof.} The conclusions \textbf{(i)}, \textbf{(ii)} and 
\textbf{(iii)} follow immediately from Lemma \ref{elementary}: for \textbf{%
(i)} each element of the sequence is of the form (\ref{abcd}); for \textbf{%
(ii)} and \textbf{(iii)} each element of the sequence is a sum of terms of
the form (\ref{abcd}).

\hfill $\Box $

\begin{remark}
The statements of Lemmas \ref{elementary} and \ref{elementary1} hold true if
the assumption of positivity is replaced by nonnegativity.
\end{remark}

The scale decomposition (\ref{vtx}) becomes effective when the Ursell
function in (\ref{betap}) is defined by a scaling limit%
\begin{equation}
\psi _{n}^{c}(\zeta _{1},\ldots ,\zeta _{n};\beta v)=\lim_{t\rightarrow
\infty }\psi _{n}^{c}(t,\zeta _{1},\ldots ,\zeta _{n};\beta v(t,\cdot ))
\label{psin}
\end{equation}%
where $\psi _{n}^{c}(t,\zeta _{1},\ldots ,\zeta _{n};\beta v(t,\cdot
))\equiv \psi ^{c}(t,\zeta _{\left\{ 1,\ldots ,n\right\} })$ is the unique
solution of the infinite system of ordinary differential equations for $%
f_{I}=f_{I}(t)\equiv f(t,\zeta _{I})$, where $\zeta _{I}=\left( \zeta
_{i_{1}},\ldots ,\zeta _{i_{k}}\right) $ is the set of variables indexed by $%
I=\left\{ i_{1},\ldots ,i_{k}\right\} \subset \left\{ 1,\ldots ,n\right\} $
and $n\in \mathbb{N}$: (see Lemma 3.3 of \cite{Brydges-Kennedy})%
\begin{equation}
\dot{f}_{I}=-\sum_{i,j\in I,~i<j}\beta \dot{v}_{ij}(t)f_{I}-\frac{1}{2}%
\sum_{J\subset I}\sum_{i\in J,~j\in I\backslash J}\beta \dot{v}%
_{ij}(t)f_{J}f_{I\backslash J}  \label{systeqs}
\end{equation}%
with (ideal gas) initial condition\footnote{%
By (\ref{vtx}), the interaction $v(t,x)$ between particles is turned off at $%
t=t_{0}$.} 
\begin{equation*}
f_{I}(t_{0})=\left\{ 
\begin{array}{ll}
1 & \text{if}\qquad \left\vert I\right\vert =1 \\ 
0 & \text{otherwise}%
\end{array}%
\right. .
\end{equation*}%
Here $\dot{v}_{ij}(t)\equiv \dot{v}(t,\zeta _{i},\zeta _{j})=\sigma
_{i}\sigma _{j}g(t)h(\left\vert x_{i}-x_{j}\right\vert /t)$ so, as $\dot{v}%
(t,\zeta _{i},\zeta _{j})$ is a measurable and translational invariant
function on the $2$--particle configuration space, $\psi ^{c}(t,\zeta _{I})$
is a measurable and translational invariant function on the $k$--particle
configuration space $\left( \left\{ -1,1\right\} \times \mathbb{R}%
^{2}\right) ^{k}$.

By the variation of constants formula the system of equations (\ref{systeqs}%
) is equivalent to a system of integrable equations: $f_{I}(t)=1$ if $%
\left\vert I\right\vert =1$ and%
\begin{equation}
f_{I}(t)=\frac{-1}{2}\int_{t_{0}}^{t}\exp \left( -\sum_{i,j\in
I,~i<j}\int_{s}^{t}\beta \dot{v}_{ij}(\tau )~d\tau \right) \sum_{J\subset
I}\sum_{i\in J,~j\in I\backslash J}\beta \dot{v}_{ij}(s)f_{J}(s)f_{I%
\backslash J}(s)~ds\ ,  \label{integralf}
\end{equation}%
if $\left\vert I\right\vert >1$, which will be usefull to our application.

\paragraph{Majorant construction for $\protect\beta <4\protect\pi $}

Using (\ref{integralf}) two of the authors have proven in \cite%
{Guidi-Marchetti} (see Theorem 2.2 and equations (4.10)-(4.12) therein) the
following

\begin{proposition}
\label{classical}Let $\Theta =\Theta (t,z)$ be the classical solution of%
\begin{equation}
\Theta _{t}=\Gamma (z^{2}\Theta ^{2})_{z}+B\left( (z\Theta )_{z}-1\right)
~,\qquad \left( t,z\right) \in \left( t_{0},\infty \right) \times \mathbb{R}%
_{+}  \label{Theta}
\end{equation}%
with $\Theta (t_{0},z)=1$ for all $z\geq 0$, where by (\ref{vtx}), (\ref{h})
and explicit calculation, $\Gamma =\Gamma (t)=\left\Vert \beta \dot{v}%
(t,\cdot )\right\Vert _{1}$ and $B=B(t)=\left\vert \beta \dot{v}%
(t,0)\right\vert /2$ are given by 
\begin{equation}
\Gamma =\beta g(t)\int_{\mathbb{R}^{2}}h(\left\vert x\right\vert /t)d^{2}x=%
\frac{\beta \pi }{4}t^{2}g(t)  \label{Gamma}
\end{equation}%
(the integral is exactly $2\pi t^{2}$ times $\displaystyle\int_{0}^{\infty }%
\dfrac{2}{\pi }\left( \arccos w-w\sqrt{1-w^{2}}\right) wdw=1/8$) and%
\begin{equation*}
B=\frac{\beta }{2}g(t)~.
\end{equation*}%
Then, the following majorant relation%
\begin{equation}
\frac{\beta }{z}\rho (\beta ,z)\leq \Theta (\infty ,z)\leq \frac{-1}{\tau
(t_{0},\infty )z}W\left( -\tau (t_{0},\infty )z\right)  \label{rho1}
\end{equation}%
holds for all $\left( \beta ,z\right) $ satisfying%
\begin{equation}
ez\tau (t_{0},\infty )<1  \label{tau1}
\end{equation}%
where%
\begin{equation}
\tau (t_{0},t)=\int_{t_{0}}^{t}\Gamma (s)\exp \left(
2\int_{s}^{t}B(s^{\prime })ds^{\prime }\right) ds~  \label{tau}
\end{equation}%
and $W(x)$ denotes the Lambert $W$--function.\cite{Corless-et-al}
\end{proposition}

\begin{remark}
\label{4pi}The proof of Proposition \ref{classical} in \cite{Guidi-Marchetti}
uses the scale decomposition (\ref{v1}) of $v$, for which $B=\beta /(4\pi t)$
and $\Gamma =2\beta t$ can be exactly calculated (for comparison, we have
set therein $\kappa (t)=1/t^{2}$ for $t\in (0,1]$). Here $v$ is given by (%
\ref{vtx}) whose scaling function $g(t)$ agree with the scaling $1/(2\pi t)$
of (\ref{v1}) only asymptotically as $t\rightarrow 0$. Writing $\tau
(t_{0},\infty )=\tau (t_{0},1)\exp \left( \beta \int_{1}^{\infty
}g(s^{\prime })ds^{\prime }\right) +\tau (1,\infty )$ together with $%
0<g(s)\leq (1+s/5)/(2\pi s)$ if $0\leq s\leq 1$ by Proposition
\ref{exponential} (see Fig \ref{fig7}), for any $0<\beta <4\pi $ the limit  
\begin{eqnarray}
\lim_{t_{0}\rightarrow 0}\tau (t_{0},1) &=&\frac{\beta \pi }{4}%
\int_{0}^{1}s^{2}g(s)\exp \left( \beta \int_{s}^{1}g(s^{\prime })ds^{\prime
}\right) ds~  \notag \\
&\leq &\frac{\beta \pi }{4}e^{\beta /10\pi }\int_{0}^{1}\frac{1}{2\pi }%
\left( s^{1-\beta /2\pi }+\frac{1}{5}s^{2-\beta /2\pi }\right) ds  \notag \\
&=&\frac{\beta \pi }{4}e^{\beta /10\pi }\left( \frac{1}{4\pi -\beta }+\frac{1%
}{5}\frac{1}{6\pi -\beta }\right)  \label{taut0}
\end{eqnarray}%
exists and $\tau (1,\infty )$ is finite since $g(t)$ decays exponentially
fast as $t\rightarrow \infty $.
\end{remark}

\begin{remark}
\label{r4pi}The existence of $\tau =\lim_{t_{0}\rightarrow 0}\tau
(t_{0},\infty )$ implies by (\ref{rho1}) and (\ref{tau1}) that the radius of
convergence $r=\sup \left\{ \left\vert z\right\vert :e\left\vert
z\right\vert \tau <1,\ z\in \mathbb{C}\right\} $ of the Mayer series (\ref%
{rho}) remains strictly positive. This fact is already remarkable
considering that $v$, given by (\ref{vtx}) with $t_{0}=0$, does not
satisfies the stability condition (\ref{stability}) (see also (\ref{Un})),
which is sufficient but not necessary for the density (\ref{rho}) be defined
in the thermodynamic limit.
\end{remark}

\noindent \textit{Proof of Proposition \ref{classical}.} By (\ref{psin}), (%
\ref{integralf}) and stability (\ref{Un}), the sequence $\left( A_{n}\right)
_{n\geq 1}$ of positive functions $A_{n}:[t_{0},\infty )\longrightarrow 
\mathbb{R}$, defined by%
\begin{equation*}
A_{n}(t)=\frac{1}{n!}\int d\varrho (\zeta _{2})\cdots d\varrho (\zeta
_{n})\left\vert \psi _{n}^{c}(\zeta _{1},\ldots ,\zeta _{n};\beta v(t,\cdot
))\right\vert
\end{equation*}%
satisfies a system of integral inequality equations%
\begin{equation}
nA_{n}(t)\leq \frac{n}{2}\int_{t_{0}}^{t}ds\exp \left(
n\int_{s}^{t}B(s^{\prime })ds^{\prime }\right) \Gamma
(s)\sum_{k=1}^{n-1}kA_{k}(s)(n-k)A_{n-k}(s)\ ,\quad n>1  \label{nAn}
\end{equation}%
with $A_{1}(t)\equiv 1$. Hence, the Mayer coefficients of the series (\ref%
{rho}) are majorized by%
\begin{equation}
n\left\vert b_{n}\right\vert \leq nA_{n}(\infty )~.  \label{bnAn}
\end{equation}

Let $\Theta (t,z)$ be defined by the series%
\begin{equation}
\Theta (t,z)=\sum_{n\geq 1}C_{n}(t)z^{n-1}=1+\sum_{n\geq 2}C_{n}(t)z^{n-1}
\label{Thetaseries}
\end{equation}%
where the sequence $\left( C_{n}\right) _{n\geq 1}$ of positive functions $%
[t_{0},\infty )\ni t\longmapsto C_{n}(t)\in \mathbb{R}_{+}$ satisfies
equations (\ref{nAn}) for $\left( nA_{n}\right) _{n\geq 1}$ as \textbf{an} 
\textbf{equality} and, consequently,%
\begin{equation}
nA_{n}(t)\leq C_{n}(t)\ ,\qquad n\geq 1\ \text{and}\ t\geq t_{0}~.
\label{nAnCn}
\end{equation}%
It can be shown (see Sec. 4 of \cite{Guidi-Marchetti}) that (\ref%
{Thetaseries}) satisfies the quasi-linear first order PDE (\ref{Theta}). So,
the first inequality of (\ref{rho1}) holds and all one needs is to determine
a domain in $\left( t_{0},\infty \right) \times \mathbb{R}_{+}$ for which
the classical solution of (\ref{Theta}) exists. Observe that (\ref{Theta})
can be written as a system of first order differential equations for the
coefficients $\left( C_{n}\right) _{n\geq 1}$. For this, by (\ref%
{Thetaseries}), we have%
\begin{eqnarray}
\Theta _{t} &=&\sum_{n\geq 1}\dot{C}_{n}z^{n-1}  \notag \\
\left( z\Theta \right) _{z} &=&\sum_{n\geq 1}nC_{n}z^{n-1}  \notag \\
(z^{2}\Theta ^{2})_{z} &=&\sum_{n\geq 2}n\left(
\sum_{k=1}^{n-1}C_{k}C_{n-k}\right) z^{n-1}~.  \label{derivatives}
\end{eqnarray}%
Substituting these series back into the equation, yields 
\begin{equation}
\dot{C}_{n}=nBC_{n}+\frac{n\Gamma }{2}\sum_{k=1}^{n-1}C_{k}C_{n-k}\ ,\qquad
n>1  \label{eqCn}
\end{equation}%
with $C_{1}(t)\equiv 1$, $t\in \lbrack t_{0},\infty )$, and initial data $%
C_{n}(t_{0})=0$ for all $n\geq 2$.

The first non-trivial equation for $n=2$,%
\begin{equation}
\dot{C}_{2}=2BC_{2}+\Gamma  \label{C2}
\end{equation}%
with $C_{2}(t_{0})=0$, has a unique solution $\tau (t_{0},t)$ given by (\ref%
{tau}), which can be written as%
\begin{equation*}
C_{2}(t)=f_{1}(t)\int_{t_{0}}^{t}\Gamma _{1}(s)ds
\end{equation*}%
where $\Gamma _{1}(s)=\Gamma (s)/f_{1}(s)$ and%
\begin{equation*}
f_{1}(t)=\exp \left( 2\int_{t_{0}}^{t}B(s^{\prime })ds^{\prime }\right)
\end{equation*}%
is an integrating factor of (\ref{C2}). As we shall see $\tau
(t_{0},t)=C_{2}(t)$ determines the radius of convergence of the series (\ref%
{Thetaseries}) for $\Theta $:%
\begin{equation}
e\left\vert z\right\vert \tau (t_{0},t)<1,  \label{eztau}
\end{equation}%
uniformly in $t_{0}$ and $t$ for $\beta <\beta _{2}$, $0<t_{0}<t<\infty $,
where $\beta _{2}=4\pi $ is the first threshold (see Remark \ref{4pi}). For
this, let $\left( C_{n}^{(1)}(t)\right) _{n\geq 1}$ be a sequence of
positive functions defined by%
\begin{equation}
\Psi (t,w)=\Theta \left( t,w/f_{1}(t)\right) =1+\sum_{n\geq
2}C_{n}^{(1)}w^{n-1}~.  \label{Psi}
\end{equation}%
Since $C_{n}^{(1)}=C_{n}/f_{1}^{n-1}$ and 
\begin{eqnarray*}
\dot{C}_{n}^{(1)} &=&\frac{\dot{C}_{n}}{f_{1}^{n-1}}-(n-1)\frac{\dot{f}_{1}}{%
f_{1}}~\frac{C_{n}}{f_{1}^{n-1}} \\
&=&\frac{\dot{C}_{n}}{f_{1}^{n-1}}-2(n-1)B~\frac{C_{n}}{f_{1}^{n-1}}
\end{eqnarray*}%
equation (\ref{eqCn}) in terms of the new $C_{n}^{(1)}$'s reads%
\begin{equation}
\dot{C}_{n}^{(1)}=-(n-2)BC_{n}^{(1)}+\frac{n\Gamma _{1}}{2}%
\sum_{k=1}^{n-1}C_{k}^{(1)}C_{n-k}^{(1)}\ ,\qquad n>1  \label{Cn1}
\end{equation}%
with $C_{1}^{(1)}(t)\equiv 1$ and initial data $C_{n}^{(1)}(t_{0})=0$ for
all $n\geq 2$. Since the coefficient $-(n-2)B$ of the linear term is
nonpositive for all $n\geq 2$, the solution of the above initial value
problem (IVP) can, in turn, be majorized by another sequence $\left( \tilde{C%
}_{n}^{(1)}\right) _{n\geq 1}$: 
\begin{equation}
C_{n}^{(1)}(t)\leq \tilde{C}_{n}^{(1)}(t)  \label{CnCn}
\end{equation}%
which solves the IVP 
\begin{equation*}
\tilde{C}_{n}^{(1)}=\frac{n\Gamma _{1}}{2}\sum_{k=1}^{n-1}\tilde{C}_{k}^{(1)}%
\tilde{C}_{n-k}^{(1)}\ ,\qquad n>1
\end{equation*}%
with $\tilde{C}_{1}^{(1)}(t)\equiv 1$ and initial data $\tilde{C}%
_{n}^{(1)}(t_{0})=0$ for all $n\geq 2$.

\medskip

\noindent \textit{Proof }of\textit{\ (\ref{CnCn}).} Using the notation
introduced in Lemma \ref{elementary1}, we write $a=\left( a_{n}\right)
_{n\geq 1}$ and $b=\left( b_{n}\right) _{n\geq 1}$ with $a_{1}=b_{1}\equiv 0$%
, $a_{n}(t)=(n-2)B(t)$ and $b_{n}(t)=n\Gamma _{1}(t)/2$ for $n>1$. The
difference sequence $\Delta =\left( \Delta _{n}\right) _{n\geq 1}$, given by 
$\Delta _{1}\equiv 0$ and $\Delta _{n}(t)=\tilde{C}%
_{n}^{(1)}(t)-C_{n}^{(1)}(t)$ for $n>1$, thus satisfies 
\begin{equation*}
\dot{\Delta}=a\cdot \Delta +b\cdot \left( \tilde{C}^{(1)}\ast \tilde{C}%
^{(1)}-C^{(1)}\ast C^{(1)}\right) \ .
\end{equation*}
Let us assume that (\ref{CnCn}) holds for some $t\geq t_{0}$. Then, by Lemma %
\ref{elementary1} we have $\dot{\Delta}(t)\geq 0$ which, together with $%
\Delta (t_{0})\equiv 0$, implies that $\Delta (t)\geq 0$. Consequently, (\ref%
{CnCn}) holds for all $t\geq t_{0}$.\hfill $\Box $

It is shown in Sec. 5 of \cite{Guidi-Marchetti} that the power series
analogous to (\ref{Psi}): $\tilde{\Psi}(t,w)=1+\displaystyle\sum_{n\geq 2}%
\tilde{C}_{n}^{(1)}w^{n-1}$ satisfies an equation given by (\ref{Theta})
setting $B=0$, $\Gamma =\Gamma _{1}$ and together with $\tilde{\psi}%
(t_{0},z)\equiv 1$ has by the method of characteristics the classical
solution 
\begin{equation*}
\tilde{\Psi}(t,w)=\frac{-1}{\tilde{\tau}_{1}(t_{0},t)w}W\left( -\tilde{\tau}%
_{1}(t_{0},t)w\right)
\end{equation*}%
provided $e\left\vert w\right\vert \tilde{\tau}_{1}(t_{0},t)<1$ holds, where 
$\tilde{\tau}_{1}(t_{0},t)=\displaystyle\int_{t_{0}}^{t}\Gamma _{1}(s)ds$
and $W(x)$ denotes the Lambert $W$--function defined implicitly by $We^{W}=x$%
, whose Taylor series about $x=0$ (see Lagrange-Bürmann theorem \cite{Davis}%
): $W(x)=\displaystyle\sum_{n\geq 1}\left( -n\right) ^{n-1}x^{n}/n!$
converges for $\left\vert x\right\vert <1/e$, including the branching point
at $x=-1/e$ (see e.g. \cite{Corless-et-al}).

Joining equations (\ref{bnAn}), (\ref{nAnCn}) and (\ref{CnCn}) together, we
conclude 
\begin{equation*}
\Theta (t,z)=\Psi (t,w)\leq \tilde{\Psi}(t,w)=\frac{-1}{\tau _{1}(t_{0},t)z}%
W\left( -\tau (t_{0},t)z\right)
\end{equation*}%
with $\tau _{1}(t_{0},t)$ given by (\ref{tau}), establishing the second
inequality of (\ref{rho1}).

\hfill $\Box $

Note that, since $f_{1}(t)>1$ for any $t_{0}$ and $t$ fixed, the radius of
convergence of the majorant series $\Theta (t,z)=\Psi (t,w)$ is smaller in $%
z $ than in $w$ variable. However, in view of Remarks \ref{4pi} and \ref%
{r4pi}, it remains strictly positive when the cutoff $t_{0}$ is removed
provided $\beta <\beta _{2}$ where $\beta _{2}=4\pi $ is the first threshold.

\paragraph{Majorant construction for $\protect\beta $ inside the threshold
intervals $I_{n}$}

The procedure of finding a majorant series for the density function can be
extended for the inverse temperature $\beta $ in the thresholds interval $%
I_{n}=[\beta _{2n},\beta _{2(n+1)})$, $n\in \mathbb{N}$ where, for
convenience, we write $\beta _{k+1}=8\pi \left( 1-1/(k+1)\right) =8\pi
k/(k+1)$. We shall present an scheme of avoiding neutral cluster collapse
which holds for any thresholds interval when the cutoff $t_{0}$ is removed.
The scheme consists of three stages. Firstly, for $\beta \in I_{(k-1)/2}$
where $k>1$ is an odd number, we remove from the system any neutral clusters
or subclusters of size less than $k$. Such a removal, which prevents the
increasing of the corresponding terms in the Mayer expansion when they
collapse, is expressed in terms of the majorant method, by inserting
Lagrange multipliers into the equation (\ref{Theta}), as many as neutral
clusters were removed.

The second stage addresses non neutral clusters of size smaller or equal to $%
k$ in the majorant equation (\ref{Theta}) which has been overestimated by
using the stability bound (\ref{stability}) instead of (\ref{stability1})
(see equation (\ref{nAn})). This issue is fixed by replacing the coefficient 
$nB$ of the linear term of (\ref{eqCn}) by $(n-1)B$. Before we apply the
second stage, we shall extract (insirting a Lagrange multiplier $L_{k}$) an
exact amount from the linear term of (\ref{eqCn}) that allows the solution
of (\ref{Theta}) for $\beta <\beta _{k+1}$ be majorized by a series with
positive radius of convergence, uniformly on cutoff $t_{0}$ (see Secs. 6 and
7 of \cite{Guidi-Marchetti}). The Lagrange multiplier $L_{k}$ of order $k$
is given by the Cesàro mean of the first $k$ Taylor series of $(z\Theta
)_{z} $ around $z=0$, truncated at order $0\leq j<k$, and this choice is
optimal.

\begin{proposition}
\label{classical1}For any $k\in \mathbb{N}$, let $\Theta =\Theta (t,z)$ be
the classical solution of%
\begin{equation}
\Theta _{t}=\Gamma (z^{2}\Theta ^{2})_{z}+B\left( (z\Theta
)_{z}-L_{k}\right) ~,\qquad \left( t,z\right) \in \left( t_{0},\infty
\right) \times \mathbb{R}_{+}  \label{Thetak}
\end{equation}%
with $\Theta (t_{0},z)=1$ for all $z\geq 0$, where $\Gamma =\Gamma
(t)=\left\Vert \beta \dot{v}(t,\cdot )\right\Vert _{1}$ and $%
B=B(t)=\left\vert \beta \dot{v}(t,0)\right\vert /2$ are given in Proposition %
\ref{classical} and%
\begin{equation*}
L_{k}=L_{k}(t)=1+\sum_{j=1}^{k-1}\left( 1-\frac{j}{k}\right) \frac{1}{j!}%
z^{j}\Theta _{\underset{j-\mathrm{times}}{\underbrace{z\cdots z}}}(t,0)
\end{equation*}%
is a Lagrange multiplier. Then, the following majorant relation%
\begin{equation}
\Theta (t,z)\leq \frac{-1}{\tau _{k}(t_{0},t)z}W\left( -\tau
_{k}(t_{0},t)z\right)  \label{ThetaW}
\end{equation}%
holds for all $\left( \beta ,z\right) $ satisfying%
\begin{equation*}
ez\tau _{k}(t_{0},\infty )<1
\end{equation*}%
where%
\begin{equation}
\tau _{k}(t_{0},t)=\int_{t_{0}}^{t}\Gamma (s)\exp \left( \frac{k+1}{k}%
\int_{s}^{t}B(s^{\prime })ds^{\prime }\right) ds~  \label{tauk}
\end{equation}%
and $W(x)$ denotes the Lambert $W$--function.\cite{Corless-et-al}
\end{proposition}

\begin{remark}
A calculation analogous to (\ref{taut0}) yields that%
\begin{eqnarray*}
\lim_{t_{0}\rightarrow 0}\tau _{k}(t_{0},1) &=&\lim_{t_{0}\rightarrow 0}%
\frac{\beta \pi }{4}\int_{t_{0}}^{1}s^{2}g(s)\exp \left( \frac{k+1}{k}\frac{%
\beta }{2}\int_{s}^{1}g(s^{\prime })ds^{\prime }\right) ds~ \\
&\leq &\frac{\beta \pi }{4}e^{2\beta /5\beta _{k+1}}\int_{t_{0}}^{1}\frac{1}{%
2\pi }\left( s^{1-2\beta /\beta _{k+1}}+\frac{1}{5}s^{2-2\beta /\beta
_{k+1}}\right) ds \\
&=&\frac{\beta }{16}e^{2\beta /5\beta _{k+1}}\left( \frac{1}{1-\beta /\beta
_{k+1}}+\frac{1}{5}\frac{1}{2-\beta /\beta _{k+1}}\right) ~,~
\end{eqnarray*}%
exists for $\beta <\beta _{k+1}$ and the radius of convergence of the
majorant series (\ref{ThetaW}) is strictly positive.
\end{remark}

\noindent \textit{Proof of Proposition \ref{classical1}.} We follow closely
the proof of Proposition \ref{classical}. Let $\Theta (t,z)$ be defined by
the series (\ref{Thetaseries}). Observe that, by%
\begin{equation*}
\left( z\Theta \right) _{z}-L_{k}=\sum_{n=2}^{k}\left( n-\frac{k-n+1}{k}%
\right) C_{n}z^{n-1}+\sum_{n\geq k+1}nC_{n}z^{n-1}
\end{equation*}%
and the remaining series of (\ref{derivatives}), (\ref{Thetak}) can be
written as a system of first order differential equations for $\left(
C_{n}\right) _{n\geq 1}$:%
\begin{equation}
\dot{C}_{n}=\frac{k+1}{k}\left( n-1\right) BC_{n}+\frac{n\Gamma }{2}%
\sum_{k=1}^{n-1}C_{k}C_{n-k}\ ,\qquad 1<n\leq k  \label{Cn1k}
\end{equation}%
and (\ref{eqCn}) for $n>k$, with $C_{1}(t)\equiv 1$ and initial data $%
C_{n}(t_{0})=0$ for $n\geq 2$. The equation for $n=2$%
\begin{equation}
\dot{C}_{2}=\frac{k+1}{k}BC_{2}+\Gamma  \label{C2k}
\end{equation}%
with $C_{2}(0)=0$ has a unique solution given by (\ref{tauk}), which can be
written as%
\begin{equation*}
C_{2}(t)=f_{k}(t)\int_{t_{0}}^{t}\Gamma _{k}(s)ds
\end{equation*}%
where $\Gamma _{k}(s)=\Gamma (s)/f_{k}(s)$ and%
\begin{equation*}
f_{k}(t)=\exp \left( \frac{k+1}{k}\int_{t_{0}}^{t}B(s^{\prime })ds^{\prime
}\right)
\end{equation*}%
is an integrating factor of (\ref{C2k}).

Let $\left( C_{n}^{(k)}(t)\right) _{n\geq 1}$ be a sequence of positive
functions defined by%
\begin{equation}
\Psi (t,w)=\Theta \left( t,w/f_{k}(t)\right) =1+\sum_{n\geq
2}C_{n}^{(k)}(t)w^{n-1}~.  \label{Psik}
\end{equation}%
Since $C_{n}^{(k)}=C_{n}/f_{k}^{n-1}$ and 
\begin{eqnarray*}
\dot{C}_{n}^{(k)} &=&\frac{\dot{C}_{n}}{f_{k}^{n-1}}-(n-1)\frac{\dot{f}_{k}}{%
f_{k}}~\frac{C_{n}}{f_{k}^{n-1}} \\
&=&\frac{\dot{C}_{n}}{f_{k}^{n-1}}-\frac{k+1}{k}(n-1)B~\frac{C_{n}}{%
f_{k}^{n-1}}~,
\end{eqnarray*}%
the equations (\ref{Cn1k}) for $1<n\leq k$ and (\ref{eqCn}) for $n>k$ in
terms of the new $C_{n}^{(k)}$'s read%
\begin{eqnarray}
\dot{C}_{n}^{(k)} &=&\frac{n\Gamma _{k}}{2}%
\sum_{j=1}^{n-1}C_{j}^{(k)}C_{n-j}^{(k)}\ ,\qquad 1<n\leq k  \notag \\
\dot{C}_{n}^{(k)} &=&-(\frac{n-k-1}{k})BC_{n}^{(k)}+\frac{n\Gamma _{k}}{2}%
\sum_{j=1}^{n-1}C_{j}^{(k)}C_{n-j}^{(k)}\ ,\qquad n>k  \label{Cnk}
\end{eqnarray}%
with $C_{1}^{(k)}(t)\equiv 1$ and initial data $C_{n}^{(k)}(t_{0})=0$ for $%
n\geq 2$. Since the coefficient $-(n-k-1)B/k$ of the linear term of (\ref%
{Cnk}) is nonpositive for all $n\geq k+1$, the solution of the above IVP can
be majorized by another sequence $\left( \tilde{C}_{n}^{(k)}\right) _{n\geq
1}$: 
\begin{equation}
C_{n}^{(k)}(t)\leq \tilde{C}_{n}^{(k)}(t)  \label{CnCnk}
\end{equation}%
which solves the IVP 
\begin{equation*}
\tilde{C}_{n}^{\prime (k)}=\frac{n\Gamma _{1}}{2}\sum_{j=1}^{n-1}\tilde{C}%
_{j}^{(k)}\tilde{C}_{n-j}^{(k)}\ ,\qquad n>1
\end{equation*}%
with $\tilde{C}_{1}^{(k)}(t)\equiv 1$ and initial data $\tilde{C}%
_{n}^{(k)}(t_{0})=0$ for $n\geq 2$. For (\ref{CnCnk}), one may apply the
same proof of (\ref{CnCn}) based in Lemma \ref{elementary1}. It follows that
(see in Sec. 7 of \cite{Guidi-Marchetti}) the power series $\tilde{\Psi}%
(t,w)=1+\displaystyle\sum_{n\geq 2}\tilde{C}_{n}^{(k)}w^{n-1}$ satisfies (%
\ref{Theta}) setting $B=0$, $\Gamma =\Gamma _{k}$ and together with $\tilde{%
\psi}(t_{0},z)\equiv 1$ has the classical solution 
\begin{equation*}
\tilde{\Psi}(t,w)=\frac{-1}{\tilde{\tau}_{k}(t_{0},t)w}W\left( -\tilde{\tau}%
_{k}(t_{0},t)w\right) 
\end{equation*}%
provided $e\left\vert w\right\vert \tilde{\tau}_{k}(t_{0},t)<1$ holds, where 
$\tilde{\tau}_{k}(t_{0},t)=\displaystyle\int_{t_{0}}^{t}\Gamma _{k}(s)ds$
and $W(x)=\displaystyle\sum_{n\geq 1}\left( -n\right) ^{n-1}x^{n}/n!$
denotes the Lambert $W$--function.

We thus have 
\begin{equation*}
\Theta (t,z)=\Psi (t,w)\leq \tilde{\Psi}(t,w)=\frac{-1}{\tau _{k}(t_{0},t)z}%
W\left( -\tau _{k}(t_{0},t)z\right) ~,
\end{equation*}%
concluding the proof of Proposition \ref{classical1}.

\hfill $\Box $

Returning to the second stage of our scheme, we show that the equation (\ref%
{Theta}), under that operation, is replaced by%
\begin{equation}
\Theta _{t}=\Gamma (z^{2}\Theta ^{2})_{z}+Bz\Theta _{z}~.  \label{Theta2}
\end{equation}%
For this, let the argument $n$ of the exponential in (\ref{nAn}) be replaced
by $n-1$. The modified coefficients $\left( C_{n}\right) _{n\geq 1}$ of the
power series (\ref{Thetaseries}) satisfy then a system of integral equations%
\begin{equation}
C_{n}(t)=\frac{n}{2}\int_{t_{0}}^{t}dse^{(n-1)\gamma (s,t)}\Gamma
(s)\sum_{k=1}^{n-1}C_{k}(s)C_{n-k}(s)\ ,\quad n>1  \label{Cn}
\end{equation}%
with $C_{1}(t)\equiv 1$ where $\gamma (s,t)=\displaystyle\int_{s}^{t}B(s^{%
\prime })ds^{\prime }$. Summing equation (\ref{Cn}) multiplied by $z^{n-1}$
over $n$ yields an integral equation for $\Theta $:%
\begin{equation}
\Theta (t,z)=1+\frac{1}{2}\int_{t_{0}}^{t}dse^{-\gamma (s,t)}\Gamma
(s)\left( z^{2}e^{2\gamma (s,t)}\Theta ^{2}(s,ze^{\gamma (s,t)})\right) _{z}
\label{Theta3}
\end{equation}%
and from this we deduce (\ref{Theta2}). Observe that an extra factor $%
e^{-\gamma (s,t)}$ inside the integration results from the stability
improvement (\ref{stability1}) and the derivative with respect to $t$
applied to this factor produces an additional term $B\left( \Theta -1\right) 
$ which has to be subtracted (due to the minus sign of the exponent) from
the last term on the right hand side of (\ref{Theta}): $B\left( (z\Theta
)_{z}-1\right) -B\left( \Theta -1\right) =Bz\Theta _{z}$.

The improved equation (\ref{Theta2}) leads to a significant outcome
regarding the radius of convergence of the Mayer series (\ref{rho}) for $%
\beta $ inside each threshold interval $I_{n}=[\beta _{2n},\beta _{2(n+1)})$%
, $n\in \mathbb{N}$.

\begin{proposition}
\label{classical3}Let $\Theta =\Theta (t,z)$ be the classical solution of (%
\ref{Theta2}) with $B$ and $\Gamma $ as in Proposition \ref{classical}. Then
the following majorant relation%
\begin{equation*}
\Theta (t,z)\leq \frac{-1}{\tau _{k}(t_{0},t)z}W\left( -\tau
_{k}(t_{0},t)z\right)
\end{equation*}%
holds for all $k\in \mathbb{N}$ and $\left( \beta ,z\right) $ satisfying $%
ez\tau _{k}(t_{0},t)<1$, where $\tau _{k}$ is given by (\ref{tauk}).
\end{proposition}

\noindent \textit{Proof.} Let $\Theta (t,z)$ be defined by the series (\ref%
{Thetaseries}) and observe that, by%
\begin{equation*}
z\Theta _{z}=\sum_{n=2}^{\infty }\left( n-1\right) C_{n}z^{n-1}~,
\end{equation*}%
(\ref{Theta2}) can be written as a system of first order differential
equations for $\left( C_{n}\right) _{n\geq 1}$:%
\begin{equation}
\dot{C}_{n}=\left( n-1\right) BC_{n}+\frac{n\Gamma }{2}%
\sum_{k=1}^{n-1}C_{k}C_{n-k}\ ,\qquad 1<n\leq k  \label{modeqCn}
\end{equation}%
with $C_{1}(t)\equiv 1$ and initial data $C_{n}(t_{0})=0$ for $n\geq 2$.
Since the coefficient $(n-1)B$ of the linear term of (\ref{modeqCn}) is
smaller than $(n-1)(k+1)B/k$ for $2\leq n\leq k$ and smaller than $nB$ for
all $n\geq k+1$,\footnote{%
This part would not be necessary for keeping positive the radius of
convergence, uniformly in $t_{0}$, at $\beta <\beta _{k+1}$.} for any $k\in 
\mathbb{N}$, the solution of the above IVP can be majorized, in view of
Lemma \ref{elementary1}, by the solution of the IVP in (\ref{Cn1k}), which
by Proposition \ref{classical1} satisfies (\ref{ThetaW}). The proof of
Proposition \ref{classical3} is concluded.

\hfill $\Box $

\paragraph{Stability of a neutral pair in the presence of other particles}

The third and last stage of our scheme deals with neutral subclusters of
order smaller than $k$ that are part of a cluster of order larger or equal
to $k+1$. So far, we have proved a weak version of the Conjecture \ref%
{conjecture}. Let $k>1$ an odd number and suppose that all neutral clusters
and subclusters of order smaller than $k$ have their singularities been
removed by hand. Then, the density function (\ref{rho}), after the removal,
satisfies%
\begin{equation*}
\frac{\beta }{z}\left\vert \rho (\beta ,z)\right\vert \leq \frac{-1}{\tau
_{k}(t_{0},\infty )z}W\left( -\tau _{k}(t_{0},\infty )z\right) ~,
\end{equation*}%
and the majorant series has strictly positive radius of convergence
uniformly in the cutoff $t_{0}$ for $\beta <\beta _{k+1}$. From the point of
view of the Mayer coefficients $b_{n}$ however, for $n\leq k$ our hypotheses
are better than the formulated in the conjecture -- instead of removing the
coefficients $b_{n}$'s entirely we remove the part of these that diverges as 
$t_{0}\rightarrow \infty $. The weakness of our hypotheses is that no
assumptions on the coefficients $b_{n}$ for $n>k$ are made in Conjecture \ref%
{conjecture}. The situations here is different from what we have done
before. According to Proposition \ref{classical1}, when $n$ is larger than $%
k $, we don't need improve the stability condition and we actually cannot
for neutral clusters of even size $n$. However, no assumptions mean that
neutral subclusters of size smaller than $k$ inside a cluster of order $n$
do not diverges as $t_{0}$ tends to $0$ and this claim needs to be proven.

To deal with this scenario, instead of a sequence $\left( A_{n}\right)
_{n\geq 1}$ defined by (\ref{nAn}), we introduce a sequence $\left( \tilde{A}%
_{m}\right) _{m\geq 1}$ of appended at $\zeta _{0}$ analogous quantities 
\begin{equation}
\tilde{A}_{m}(s,\sigma _{1},\ldots ,\sigma _{m})=\frac{1}{m!}\int_{\mathbb{R}%
^{2}\times \cdots \times \mathbb{R}^{2}}dx_{1}\cdots dx_{m}\left\vert
\sum_{j=1}^{m}\sigma _{0}g(s)h(\left\vert x_{j}-x_{0}\right\vert /s)\sigma
_{j}\psi _{m}^{c}(s,\zeta _{1},\ldots ,\zeta _{m})\right\vert
\label{Atildem}
\end{equation}%
which are independent of $\zeta _{0}=(x_{0},\sigma _{0})$ by translational
invariance of variable $x_{0}$ and $\left\vert \sigma _{0}\right\vert =1$.
In the next paper we shall study in particular the recursion relations
satisfied by theses quantities together with their majorant equations as
well as a systematic majorant approach for the correlation function. In the
present paper, we shall restrict ourselves to the simplest case of $m=2$ of (%
\ref{Atildem}).

Let us consider a $n$--particle cluster containing a neutral pair
subcluster. Referring to the formula (\ref{integralf}), let $I$ be an index
set of a cluster with $\left\vert I\right\vert =n$ and let $J$ be the index
set of a pair $\left\vert J\right\vert =2$ of particles with opposite
charges: $\sigma _{1}\sigma _{2}=-1$ located at $x_{1}$ and $x_{2}$, whose
Ursell function at scale $s$ is simply given by%
\begin{equation}
\psi _{2}^{c}(s,\zeta _{1},\zeta _{2})=\beta \int_{t_{0}}^{s}g(\tilde{s})h(r/%
\tilde{s})\exp \left( \beta \int_{\tilde{s}}^{s}g(\tau )h(r/\tau )d\tau
\right) d\tilde{s}~.  \label{psi2}
\end{equation}%
where $r=\left\vert x_{2}-x_{1}\right\vert $. We refer to Sec. 6.3 of \cite%
{Guidi-Marchetti} for detail. Let $\Delta =\Delta (s,\tilde{s}%
,x_{0},x_{1},x_{2})$ be the $h$ part of (\ref{Atildem}) including (\ref{Psi}%
) given by 
\begin{equation}
\Delta =\left( h(\left\vert x_{0}-x_{1}\right\vert /s)-h(\left\vert
x_{0}-x_{2}\right\vert /s)\right) h(\left\vert x_{1}-x_{2}\right\vert /%
\tilde{s})~  \label{Deltahhh}
\end{equation}%
with $t_{0}\leq \tilde{s}<s$. Using the convolution form (\ref{s2h}) of the
Euclid's hat $h(w)$ together with its geometric interpretation as the area
of \textquotedblleft caps\textquotedblright\ (see proof of Proposition \ref%
{euclidhat}), we shall find and upper bound for the integral over $x_{1}$
and $x_{2}$ of this quantity. For this, we write 
\begin{equation*}
\Delta =\frac{4}{\pi s^{2}}\int_{\mathbb{R}^{2}}dz\cdot \frac{4}{\pi \tilde{s%
}^{2}}\int_{\mathbb{R}^{2}}d\tilde{z}\chi _{s/2}(x_{0}-z)\left( \chi
_{s/2}(z-x_{1})-\chi _{s/2}(z-x_{2})\right) \chi _{\tilde{s}/2}(x_{1}-\tilde{%
z})\chi _{\tilde{s}/2}(\tilde{z}-x_{2})~
\end{equation*}%
and observe that the integrand of $\Delta $ differs from $0$ if, and only
if, either $x_{1}$ is inside of the non null intersection $B_{s/2}(z)\cap B_{%
\tilde{s}/2}(\tilde{z})$ and $x_{2}$ is inside the complementary region $B_{%
\tilde{s}/2}(\tilde{z})\backslash (B_{s/2}(z)\cap B_{\tilde{s}/2}(\tilde{z}%
))\neq \emptyset $ or vice-versa. As a consequence, we have%
\begin{equation*}
\int_{\mathbb{R}^{2}\times \mathbb{R}^{2}}dx_{1}dx_{2}\left\vert \chi
_{s/2}(z-x_{1})-\chi _{s/2}(z-x_{2})\right\vert \chi _{\tilde{s}/2}(x_{1}-%
\tilde{z})\chi _{\tilde{s}/2}(\tilde{z}-x_{2})=2A\cdot B
\end{equation*}%
where, denoting by $\left\vert D\right\vert $ the area of a bounded domain $%
D\subset \mathbb{R}^{2}$, $A=\left\vert B_{s/2}(z)\cap B_{\tilde{s}/2}(%
\tilde{z})\right\vert $ and $B=\left\vert B_{\tilde{s}/2}(\tilde{z}%
)\backslash (B_{s/2}(z)\cap B_{\tilde{s}/2}(\tilde{z}))\right\vert =\left(
\pi \tilde{s}^{2}/4\right) -A$. Using%
\begin{equation*}
2A\cdot B=\frac{1}{2}\left( A+B\right) ^{2}-\frac{1}{2}\left( A-B\right)
^{2}\leq \frac{1}{2}\left( A+B\right) ^{2}=\frac{1}{2}\left( \frac{\pi 
\tilde{s}^{2}}{4}\right) ^{2}
\end{equation*}%
and the fact that $A$ and $B$ are different from $0$ if and only if%
\begin{equation*}
\frac{s-\tilde{s}}{2}<\left\vert z-\tilde{z}\right\vert <\frac{s+\tilde{s}}{2%
}
\end{equation*}%
we have%
\begin{eqnarray}
\int_{\mathbb{R}^{2}\times \mathbb{R}^{2}}dx_{1}dx_{2}\left\vert \Delta (s,%
\tilde{s},x_{0},x_{1},x_{2})\right\vert  &=&\frac{4}{\pi s^{2}}\int_{\mathbb{%
R}^{2}}dz~\chi _{s/2}(x_{0}-z)\cdot \frac{4}{\pi \tilde{s}^{2}}\int_{\mathbb{%
R}^{2}}d\tilde{z}~2A\cdot B  \notag \\
&\leq &\frac{1}{\tilde{s}^{2}}\left( \left( s+\tilde{s}\right) ^{2}-\left( s-%
\tilde{s}\right) ^{2}\right) \cdot \frac{1}{2}\left( \frac{\pi \tilde{s}^{2}%
}{4}\right) ^{2}  \notag \\
&=&\frac{1}{8}\pi ^{2}\tilde{s}^{3}s~,  \label{ss}
\end{eqnarray}%
and this implies that $\tilde{A}_{2}(s)$ given by (\ref{Atildem}) with $m=2$
and $\sigma _{1}\sigma _{2}=-1$ is bounded uniformly with respect to the
cutoff $t_{0}$ provided $\beta \in \lbrack 0,6\pi )$, i. e., inside the
first threshold interval $I_{1}=[4\pi ,6\pi )$. For this, observe that by
Proposition \ref{exponential} (see (\ref{taut0}))%
\begin{eqnarray}
\tilde{A}_{2}(s) &=&\frac{1}{2}g(s)\int_{\mathbb{R}^{2}\times \mathbb{R}%
^{2}}dx_{1}dx_{2}\beta \int_{t_{0}}^{s}g(\tilde{s})\left\vert \Delta (s,%
\tilde{s},x_{0},x_{1},x_{2})\right\vert \exp \left( \beta \int_{\tilde{s}%
}^{s}g(\tau )h(\left\vert x_{1}-x_{2}\right\vert /\tau )d\tau \right) d%
\tilde{s}  \notag \\
&\leq &\frac{\beta }{64}m(s)\int_{t_{0}}^{s}\tilde{s}^{2}m(\tilde{s})\exp
\left( \frac{\beta }{2\pi }\int_{\tilde{s}}^{s}\frac{1}{\tau }m(\tau )d\tau
\right) d\tilde{s}  \notag \\
&<&C\int_{t_{0}}^{1}\tilde{s}^{2-\beta /2\pi }d\tilde{s}<\frac{C}{3-\beta
/2\pi }<\infty   \label{A2tilde}
\end{eqnarray}%
if $\beta <6\pi $, uniformly in $t_{0}$.

The integral of (\ref{Deltahhh}) performed over $x_{1}$ and $x_{2}$
desregarding the minus sign would be proportional to $\tilde{s}^{2}s^{2}$,
by (\ref{Gamma}). The small cluster neutrality condition leads to a
rearrangement of the powers in $\tilde{s}$ and $s$ (\ref{ss}) in favor of $%
\tilde{s}$. Note that $sg(s)$ remains integrable by Proposition \ref%
{exponential} and a new function $\Gamma (s)$ needs to be redefined
accordingly. The rearrangement is not enough to prevent the collapse of the
neutral pair inside the high order threshold intervals and we need to be
more careful when $\left\vert x_{1}-x_{2}\right\vert /\tau $ is small. By
the first mean value theorem, there exist $\tau ^{\ast }\in \left[ \tilde{s}%
,s\right] $ such that%
\begin{equation*}
\frac{\beta }{2\pi }\int_{\tilde{s}}^{s}m(\tau )h(\left\vert
x_{1}-x_{2}\right\vert /\tau )\frac{d\tau }{\tau }=m(\tau ^{\ast
})h(\left\vert x_{1}-x_{2}\right\vert /\tau ^{\ast })\frac{\beta }{2\pi }%
\log \frac{s}{\tilde{s}}~.
\end{equation*}%
For fixed $s$, let us say $s=1$, let $\Lambda =\left\{ \left( r,\tilde{s}%
\right) \in \mathbb{R}_{+}\times \left[ t_{0},1\right] :r/\tau ^{\ast }(r,%
\tilde{s})\leq 0.2\right\} $ and note that $m(\tau ^{\ast })h(r/\tau ^{\ast
})<3/4$ for $\left( r,\tilde{s}\right) $ in the complementary set $(\mathbb{R%
}_{+}\times \left[ t_{0},1\right] )\backslash \Lambda $, by Propositions \ref%
{euclidhat} and \ref{exponential}. Under this condition $\tilde{A}_{2}(s)$
can be bounded by the last integral in (\ref{A2tilde}) with the exponent $%
2-\beta /2\pi $ of $\tilde{s}$ replaced by $2-(3/4)\beta /2\pi =2-3\beta
/8\pi $, which is finite for $\beta <8\pi $. On the other hand, one can show
that the integral of $\Delta $ over $\left( x_{1},x_{2}\right) \in \mathbb{R}%
^{2}\times \mathbb{R}^{2}$ in (\ref{ss}) restricted to $\Lambda $, where $%
r=\left\vert x_{1}-x_{2}\right\vert $, is proportional to $\tilde{s}^{4}$
(instead of $\tilde{s}^{3}$) under the change of variables $x_{i}=\tilde{s}%
y_{i}$, $i=1,2$. Observe that $\tau ^{\ast }(r,\tilde{s})$ tends to $\tilde{s%
}$ when $r$ tends to $0$. As a consequence, $\tilde{A}_{2}(s)$ can be
bounded by the last integral in (\ref{A2tilde}) with the exponent $2-\beta
/2\pi $ of $\tilde{s}$ replaced by $3-\beta /2\pi $ which is finite again
for $\beta <8\pi $.

\section{Summary and open question}

The main result of the present paper, Theorem \ref{main}, states that the
energy $U_{n}(\xi ;h)$ of a configuration $\xi =\left( x,\sigma \right)
=\left( x_{1},\ldots ,x_{n},\sigma _{1},\ldots ,\sigma _{n}\right) $ of $n$
particles, with $\left( x_{i},\sigma _{i}\right) \in \mathbb{R}^{2}\times
\{+1,-1\}$, interacting through the two--dimensional Euclid's hat pair
potential $h\left( \cdot /s\right) $ at scale $s$ satisfies (\ref{Unh}).
Since the inequality saturates when the $n$ particles collapses all together
to a single point with net charge $0$ if $n$ is even and $\pm 1$ if $n$ is
odd, a corollary to this (see Corollary \ref{minimal}) is that the minimal
specific energy $e(h)$ and the minimal constrained specific energy $\bar{e}%
(h)$, defined by (\ref{e}) and (\ref{ebar}), are both $-1/2$. The same
statement holds to positive radial potentials of positive type in any
dimension $d\geq 2$ provided it can be written as scale mixtures
of Euclid's hat: $v(x)=\displaystyle\int g(s)h\left( \left\vert x\right\vert
/s\right) ds$, $g(s)\geq 0$ and the right hand side of (\ref{Unh}) is
multiplied by $\dint g(s)ds$. Consequently, if $n$ is odd the stability
bound (\ref{stability}) can be replaced by (\ref{stability1}) for any
potential of this class with $B=\bar{B}=\dfrac{1}{2}\displaystyle\int g(s)ds$.

We have applied the main result to the two--dimensional Yukawa gas with
particles activity $z$ at the inverse temperature $\beta $ in the interval
of collapse $[4\pi ,8\pi )$. A Cauchy majorant, proposed in
\cite{Guidi-Marchetti} for the pressure and density function, can be
written in terms of the principal branch of the $W$--Lambert function
which is analytic provided its argument $-z\tau _{k}$, with $\tau
_{k}=\tau _{k}(t_{0},t)$ given by (\ref{tauk}), satisfies $e\left\vert
  z\right\vert \tau _{k}<1$, $\beta <\beta _{2n}=8\pi \left(
  1-1/2n\right) $ when the divergent part of 
the leading even Mayer coefficients up to order $2n\leq k+1$, $k>1$, are
extracted. It has been assumed in addition that an improved stability
condition (see Conjecture 2.3 of \cite{Guidi-Marchetti}) holds for any odd
number of particles $2n-1\leq k$. However, the numerical evaluation (see
Remark 7.5 of \cite{Guidi-Marchetti}) of the total energy $U_{2n-1}(\xi
;\beta \dot{v})$ for the standard scaling decomposition (\ref{v1}) when $n=2$
and $3$ have indicated that it would fail for sufficient large $k$ and
Proposition \ref{bare3} now proves that $U_{3}(\xi ;\beta \dot{v})$ does not
satisfy the improved stability for $k>15$. We have in the present paper
proved that when the Yukawa potential $v$ is represented as scale mixtures
of Euclid's hat it satisfies Conjecture 2.3 of \cite{Guidi-Marchetti} for
any $k>1$ and, moreover, all the estimates necessary to establish
convergence of the majorant series in \cite{Guidi-Marchetti} holds for this
representation of $v$ due to Proposition \ref{exponential}. We have
reestated Propositions \ref{classical}, \ref{classical1} and \ref{classical3}
accordingly for the reader convenience.

It is important to stress at this point that the classical solution $\Theta
_{k}=\Theta _{k}(t,z)$ of (\ref{Thetak}) is actually a majorant for the
density function (\ref{rho}) and the same statement (\ref{rho}) holds in
Proposition \ref{classical1} as long as $t_{0}>0$. One open question is
whether the majorant $\Theta _{k}$ remain faithful when the cutoff $t_{0}$
is set to $0$. We answer the question afirmatively only for $k=3$ and argue
that this question might be dealt using the superstability of the
(two--species) Yukawa potential restricted to configurations in which
a neutral subcluster is located in a small volume of linear size
$t_{0}>0$ (see \cite{Gielerak, Rebenko-Tertychnyi} and references therein).

\end{document}